\documentclass{JHEP3}
\preprint{MAN/HEP/2008/10}

\usepackage{cite}
\usepackage{epsfig}
\usepackage{color}
\usepackage{rotating}
\usepackage{lscape}
\let\normalcolor\relax
\usepackage{graphics}
\usepackage{inputenc}
\usepackage{xspace}
\inputencoding{latin1}
 \renewcommand\email[1]{{\scriptsize\tt\href{mailto:#1}{#1}}}

\newcommand{\M}{\ensuremath{\mathbf{M}}}
\newcommand{\G}{\ensuremath{\mathbf{\Gamma}}}
\newcommand{\Sv}{\ensuremath{\mathbf{S}}}

\newcommand{\qbar}{\ensuremath{\overline{q}}}

\def\beq{\begin{equation}}
\def\eeq{\end{equation}}
\def\beqa{\begin{eqnarray}}
\def\eeqa{\end{eqnarray}}

\newcommand{\eqref}[1]{Eq.~(\ref{#1})\xspace}

\skip\footins = 1\bigskipamount plus 2pt minus 4pt                              
\title{\boldmath Color evolution of 2 $\rightarrow$ 3 processes}

\author{Malin Sjödahl\\
School of Physics \& Astronomy, University of Manchester, \\
  Oxford Road, Manchester M13 9PL, U.K.\\
  E-mail: \email{malin.sjodahl@manchester.ac.uk}
}
  
  \abstract{
The color structure needed for resummation of all colored $2 \to 3$
processes is calculated using multiplet inspired
$s$-channel bases. In this way the resulting matrices, describing the 
color structure, are guaranteed to obey simplifying symmetries.

  }

\begin{document}
 
 
\section{Introduction}
\label{sec:intro}

In perturbation theory resummation becomes necessary when large logarithms 
compensate the smallness of the coupling constant and invalidate a fixed 
order calculation.

Large logarithms occur in the collinear region, where a large rapidity 
logarithm gives an effectively large phase space for radiation. 
In this case the radiation is coherent and acts as if it was emitted by a 
single parton.
Large logarithms may also occur for wide angle radiation of soft gluons.
For such radiation, where real and virtual gluon emissions cancel 
(at least for global observables, where there is no issue with out 
of the gap region gluons being prevented from radiating into the gap), 
the complication of the color structure can be expressed in 
terms of a matrix containing color and phase space information: the 
soft anomalous dimension matrix 
\cite{Contopanagos:1996nh,Sotiropoulos:1993rd,Kidonakis:1998nf, Oderda:1999kr}.
This matrix has contributions both from the region of phase space where 
the exchanged gluons go on shell, sometimes referred to as the eikonal gluon
contribution, and the region of phase space where the hard partons go 
on shell at intermediate steps, the Coulomb gluon region. 
The latter would in an Abelian theory give just an unobservable phase, 
but leads to a physically relevant change for a non-Abelian theory.

In principle, soft gluon resummation of this type is of relevance for 
any colored emission where there is a large logarithm 
in (hard scale/soft scale).
However, largely due to the 
complicated color structure, the soft anomalous dimension matrix has so far 
only been calculated for the $2 \to 2$ processes 
\cite{Contopanagos:1996nh,Sotiropoulos:1993rd,Kidonakis:1998nf, Oderda:1999kr,Appleby:2003hp,Dokshitzer:2005ig,Banfi:2004yd} and one $2\to3$ process 
\cite{Kyrieleis:2005dt}.

What is used in event generators is instead a color diagonal, leading $N$,
approximation. This may work well for most inclusive observables,
but is destined to break down under certain circumstances.
Perhaps the simplest example of this type of observable
is an observable with a rapidity gap, free from radiation above some 
veto transverse momentum $Q_0$, such that $\log(\mbox{hard scale}/Q_0)$
is sufficiently large \cite{Oderda:1998en}. 
In this case color mixing effects are significant. 
In particular a colored exchange may change to an overall color singlet 
exchange by virtual gluon exchange, leading to a 
decreased radiation probability and an increased gap survival probability
\cite{Appleby:2003sj,Forshaw:2007vb}.
The results for the gap-survival probability within a rapidity region
$Y$ have also been experimentally tested using events with 
rapidity gaps at HERA and the Tevatron \cite{Derrick:1995pb,Abe:1997ie,Abe:1998ip,Abbott:1998jb,Adloff:2002em,Appleby:2003sj}.

To go beyond events with 2 jets, clearly the color evolution for 
$2 \to n$ matrix elements is needed for a $n$-jet event. However, 
due to the non-global nature of most observables, corrections from 
the color structure corresponding to $2 \to n$ scattering, for all $n$, 
are in principle present even for two jet events with a rapidity gap 
\cite{Dasgupta:2001sh, Dasgupta:2002bw, Berger:2001ns, Delenda:2006nf,Appleby:2002ke,Dokshitzer:2003uw,Berger:2003iw}.
To deal with this problem in its full complexity would probably require 
significant progress in our understanding of the group theory aspects of QCD.
What is presented here are the results needed to calculate the soft anomalous 
dimension matrix for all $2 \to 3$ processes, 
completing the work initiated in \cite{Kyrieleis:2005dt}.
This would be the full story for a global observable with a large logarithm
in (hard scale)/(veto scale) and a moderate $Y$. However, since most
observables of interest are non-global, these results can alternatively be used
to obtain the lowest order non-global correction for two jet events with a 
rapidity gap.
 
Apart from the aforementioned applications, this color structure information is also
needed to investigate the effects of the so called super-leading logarithms, 
carrying an extra power of 
$\log(\mbox{hard scale}/\mbox{veto scale})$ \cite{Forshaw:2006fk},
and to see if the possible breakdown of factorization suggested in 
\cite{Kyrieleis:2006fh} 
for $qq \to qqg$ persists to more complicated color states (as is anticipated).

The outline of this paper is as follows: 
First, to set the scene, gluon resummation is discussed in general in 
section \ref{sec:GR}, and exemplified with several $2 \to 2 $ processes.
Then, in section \ref{sec:2To3} the construction of multiplet inspired bases
is presented for $2 \to 3$ processes. The resulting soft anomalous dimension
matrices are discussed in section \ref{sec:Results}, and the explicit formulae 
are stated in appendix \ref{sec:AlgebraResults}. 
Finally some concluding remarks are made in section \ref{sec:Conclusions}.

\section{Gluon resummation}
\label{sec:GR}

To perform the resummation of virtual gluon exchange on a hard
scattering matrix element $M$, the effect of virtual 
gluon exchange on $M$ has to be considered. 
This is perhaps most easily illustrated by
a simple example. Consider therefore a hard scattering of two quarks
$q_a q_b \rightarrow q_c q_d$. The \textit{exchange} between the
quarks $q_a$ and $q_b$ can form a total color singlet 
(from for example the exchange of an electro-weak boson) 
or a total color octet (as from exchange of a gluon). This can be described 
by the color tensors 

\begin{eqnarray}
  C^1_{abcd}&=&\delta_{ac} \delta_{bd},\nonumber \\
  C^8_{abcd}&=&t^g_{ca} t^g_{db}
=\frac{1}{2}(\delta_{ad}\delta_{bc}-\frac{1}{N}\delta_{ac}\delta_{bd}).
\label{eq:qqBasis}
\end{eqnarray}
Similarly, the color structure corresponding to anti-quarks and gluons in the 
initial and final states, or more than 2+2 partons can be described.
For this purpose, in general, delta functions in quark indices, delta functions
in gluon indices, generators $t^a_{bc}$, and
the symmetric and antisymmetric structure constants $d_{abc}$ and $i f_{abc}$,
\begin{equation}
(if/d)_{abc}=2(\mbox{Tr}[t^a t^b t^c] (-/+) \mbox{Tr}[t^b t^a t^c]).
\label{eq:fd}
\end{equation}
may be used.

The color states used for describing a colored amplitude form a 
vector space. Using the scalar product

\begin{equation}
\sum_{a,b,c,d...}C^i_{abcd...}(C^j)^{*}_{abcd...},
\label{eq:SP}
\end{equation}
where $C^i$ denotes a certain color structure  and $a...d...$ are quark, 
anti-quark and gluon indices,
a complete orthogonal set of color states can be found. The effect of 
exchanging a virtual gluon on a color state is to map the color vector into 
a linear combination of color states. This linear
transformation can thus be described by a matrix.

\subsection{2 $\rightarrow$ 2 processes}

For a non-trivial color structure to arise there must be at least four
colored particles \cite{Dokshitzer:2005ig}. 
In order to set the scene for the more complicated processes, and 
shed some light on the construction of a basis, some 
$2 \to 2$ processes are discussed below.

\subsection{A simple example, $qq \rightarrow qq$}

Here a simple non-trivial case is considered, namely
$q_a q_b \to q_c q_d$ \cite{Kidonakis:1998nf}.
For the resummation to be successful (even in the Abelian case) we need to 
assume that only gluons strongly ordered in transverse momenta contribute. 
This assumption simplifies both the kinematics, as the softer momenta
can be ignored, and the color algebra, as no nestled color structures have to 
be considered. 
To deal with resummation of soft gluons, the exchange 
of \textit{any} number of virtual gluons has to be considered. 
The first step is naturally to consider the exchange 
of only one extra gluon, attached in all possible ways to the hard scattering.
For example, a virtual gluon could be exchanged in the $t$-channel. This 
would change an initial $t$-channel color singlet exchange into a color octet 
exchange.
An initial color octet exchange would transform into a combination of 
a color singlet exchange and a color octet exchange. The exchange of a 
$u$-channel gluon would have a similar effect, whereas a singlet would
stay a singlet, and an octet would stay an octet, 
under a vertex correction exchange.
The color structure further simplifies, as the color effect of an exchange 
between quark $a$ and quark $b$ is the same as the effect of an 
exchange between $c$ and $d$, and similarly for $ac$ and $bd$ or $ad$ and $bc$. 
The kinematic factors from phase space integration 
(azimuth and rapidity) over the
regions in which radiation is forbidden can thus be summed as 
$T=\Omega_{ab}+\Omega_{cd}$,
$U=\Omega_{ad}+\Omega_{bc}$ and 
$V=\Omega_{ac}+\Omega_{bd}$.
All this information is described by the soft anomalous 
dimension matrix,

\begin{equation}
  \label{eq:Gamma}
  \G_{qq \to qq} =
  \left(
  \begin{array}{ll}
    \frac{\left(N^2-1\right) V}{2 N} & \frac{\left(N^2-1\right) (T+U)}{4 N^2} \\
    T+U & -\frac{-U N^2+2 T+2 U+V}{2 N}
  \end{array}
\right),
\end{equation}
where the phase space integrals over the azimuth and rapidity of the exchanged
$k'$ are
\begin{equation}
\Omega_{ij}=
\frac{1}{2} 
s_{ij} 
\left[
\int_{\Omega} \frac{dy' d \phi'}{2 \pi} 
\frac{{k'}_{\perp}^2 p_i \cdot p_j}{2 p_i \cdot  k' k'\cdot p_j}
-\frac{1}{2}(1-s_{ij})i \pi
\right]
\label{eq:PhaseSpace}
\end{equation}
in the relevant eikonal limit. 
Here $s_{ij}=-1$ if the quarks (partons in general) $i$ and $j$ are 
both incoming or both outgoing, and $s_{ij}=1$ otherwise.
(For each involved anti-quark and each outgoing gluon - with the present 
triple gluon convention, see \eqref{eq:gggSign} -
there is an additional overall minus sign.) 
These integrals have been evaluated for a topology in which radiation
is forbidden within a rapidity region $-Y/2<y'<Y/2$ in 
for example \cite{Kyrieleis:2005dt}.
With the above simplification all virtual gluons can be resummed as

\begin{equation}
  \label{eq:M}
  \M=\exp\left(  -\frac{2}{\pi}
	 {\displaystyle\int\limits_{Q_{0}}^{Q}}
	 \alpha_{s}({k'}_{\perp}) \frac{d{k'}_{\perp}}{{k'}_{\perp}} \G \right)  \M_{0},
\end{equation}
where $\M_{0}$ denotes the hard matrix element (as a vector in color space).

The $qq \to qq$ basis \eqref{eq:qqBasis} is orthogonal but not 
normalized.
To account for this, when calculating the cross section the scalar
product of the color vectors has to be used, 
$\sigma= \M^\dagger \Sv_{qq \to qq} \M$ with

\begin{equation}
\Sv_{qq \to qq}=\left(
\begin{array}{ll}
 N^2 & 0 \\
 0 & \frac{1}{4} \left(N^2-1\right)
\end{array}
\right).
\end{equation}
As this is obtained by summing (rather than averaging) over incoming quarks
a quark color averaging is expected for the hard scattering matrix.
Similar normalization matrices will be needed for the other bases under 
consideration here.

As for the color algebra part, it should be noted that the $qq \rightarrow qq$
scattering, here described by considering the exchange in $t$-channel,
corresponds precisely to the $q \qbar \rightarrow q\qbar$ scattering,
but viewed in the $s$-channel. The result, \eqref{eq:Gamma} can be used 
for any amplitude with four external quarks or anti-quarks. 
In principle, when the color algebra has been 
calculated in one case, it can be used also for the other. 
For this reason, to describe all possible four-parton colored scattering 
processes, only three different color topologies are needed. The two remaining
are briefly described below.
Identical arguments reduce the number of different subprocesses that need to 
be considered for $2 \to 3$ processes to three:
four external quarks or anti-quarks and one gluon, 
two external quarks or anti-quarks and three gluons, 
or five external gluons.

Generally speaking a certain basis may be physically more illuminating. For example,
using the $t$-channel basis as above, one can easily make a comparison to
classical radiation from an accelerated (color) charge. 
The classical radiation 
is expected to be large when the acceleration is large. In the case of 
acceleration of colors, this corresponds to a $t$-channel octet (gluon)
exchange, and indeed, one finds that the amount of radiation is significantly 
higher for a color octet exchange than for a color singlet exchange 
\cite{Forshaw:2007vb}.

\subsection{Color structure of $qg \rightarrow qg$}

Colorwise, one may choose to describe this process in the $s$-channel basis,
stating that a quark (3) and a gluon (8) can be in three different states
$8 \otimes 3 = 3 \oplus \bar{6} \oplus 15$. 
The final quark and gluon must then be in the 
same state. One can thus write down 3 color tensors corresponding to 
$3 \rightarrow 3$, $\bar{6} \rightarrow \bar{6}$ and $15 \rightarrow 15$.

Alternatively, one can consider the exchange in the $t$-channel basis 
and conclude that the incoming quark and outgoing quark, 
counting as an incoming anti-quark, can be in a color singlet or a color 
octet states. The gluons must then also be in an octet or a singlet state 
respectively, but $8 \otimes 8$ contains two octet to choose from.
These octets may, without loss of generality, be
divided into one piece which is symmetric in the incoming gluons and one 
piece which is antisymmetric.
Alternatively one can thus use the basis corresponding to
$1 \rightarrow 1$, $8 \rightarrow 8^s$ and $8 \rightarrow 8^a$.

A basis used for $qg \rightarrow qg$ can clearly also be used to describe
$gg \rightarrow q \qbar$,
$\qbar \, \qbar \rightarrow gg$ and
$\qbar g \rightarrow \qbar g$.

\subsection{Color structure of $gg \to gg$}
\label{sec:ggTogg}

Writing down the possible multiplets of two gluons gives, 
in $SU(3)$-inspired notation:

\begin{equation}
8 \otimes  8 = 1 \oplus 8 \oplus 8 \oplus 10 \oplus \overline{10} \oplus 27 \oplus 0. 
\label{eq:8x8}
\end{equation}
Clearly the number of states are different from 8, 10 etc when $N \neq 3$,
but the above notation will nevertheless be used throughout this paper.
For three colors, the last state has multiplicity 0, 
and need thus not be considered.
(For $N=2$ there are only three states on the right hand side of \eqref{eq:8x8}
``1'', ``8'' and ``27'', with multiplicity 1, 3 and 5 respectively.)
The incoming gluons may 
be in any of the above multiplets and the outgoing must be in the same.
This enables the construction of $9$ different color states.
A basis corresponding to the 9 possible transitions is
\cite{Macfarlane}:

\begin{eqnarray}
  P^{1}_{abcd}&=&\frac{\delta _{{ab}} \delta _{{cd}}}{N^2-1}\nonumber \\
  P^{8ss}_{abcd}&=&\frac{N d_{{abg}} d_{{cdg}}}{N^2-4}\nonumber\\
  P^{8aa}_{abcd}&=&\frac{f_{{abg}} f_{{cdg}}}{N}\nonumber\\
  P^{10+\overline{10}}_{abcd}&=&\frac{1}{2} \left(\delta _{{ac}} \delta _{{bd}}-\delta _{{ad}}
   \delta _{{cb}}\right)-\frac{f_{{abg}} f_{{cdg}}}{N}\nonumber\\
   P^{27}_{abcd}&=&\frac{N d_{{abg}} d_{{cdg}}}{4(N+2)}+\frac{1}{2} f_{{adg}}
   f_{{cbg}}-\frac{1}{4} f_{{ab g}}
   f_{{cdg}}+\frac{1}{4} \delta _{{ad}} \delta
   _{{bc}}\nonumber\\
   & &+\frac{1}{4} \delta _{{ac}} \delta
   _{{bd}}+\frac{\delta _{{ab}} \delta _{{cd}}}{2 (N+1)}\nonumber\\
   P^{0}_{abcd}&=&-\frac{N d_{{ab g}} d_{{cdg}}}{4(N-2)}-\frac{1}{2} f_{{ad g}}
   f_{{cbg}}+\frac{1}{4} f_{{abg}}
   f_{{cdg}} \nonumber\\
   & &+\frac{1}{4}\delta _{{ac}} \delta
   _{{bd}}+\frac{1}{4} \delta _{{ad}} \delta _{{cb}}
   -\frac{\delta _{{ab}} \delta _{{cd}}}{2(N-1)}\nonumber\\
   P^{8as}_{abcd}&=&d_{{cdg}} i f_{{abg}}+d_{{abg}} i f_{{cdg}}\nonumber\\
   P^{8sa}_{abcd}&=&d_{{cdg}} i f_{{abg}}-d_{{abg}} i f_{{cdg}}\nonumber\\
      P^{10-\overline{10}}_{abcd}&=&\frac{1}{2} d_{{acg}} i f_{{bgd}}-\frac{1}{2} d_{{bgd}}i f_{{acg}}.
  \label{eq:ggBasis}
\end{eqnarray}
Of these color tensors,
$P^1$, $P^{8ss}$, $P^{8aa}$, $P^{10+\overline{10}}$, $P^{27}$ and $P^0$
are projectors, i.e. they satisfy
$P^i_{ABmn}P^j_{mnCD}=\delta_{ij}P^i_{ABCD}$ (no sum over $i$). 
This also defines their normalizations.
For three colors $P^0$ annihilates anything.
The last three tensors are not projectors,
as can be seen from the fact that they have different symmetries w.r.t. 
interchanging the index pairs $ab$ and $cd$. However $P^{10-\overline{10}}$ 
still annihilates any non-decuplet or anti-decuplet state, 
and $P^{8as}$ and $P^{8sa}$ annihilate any non-octet state.
The reason for using $10+\overline{10}$ and $10-\overline{10}$ is that 
$10+\overline{10}$ is symmetric w.r.t. interchanging quarks and anti-quarks
in the root diagram.

The soft anomalous dimension matrix, in the basis \eqref{eq:ggBasis} is block 
diagonal. The color evolution matrix mixing the first 6 basis states 
(5 for $N=3$) can be stated 

\begin{eqnarray}
& &\G^{6 \times 6}_{gg \to gg}=\\
& &\left(
\begin{array}{llllll}
 -N T & 0 & N (U-V) & 0 & 0 & 0 \\
 0 & -\frac{1}{4} N (2 T+U+V) & \frac{1}{4} N (U-V) & \frac{1}{2} N (U-V) & 0 & 0 \\
 \frac{N (U-V)}{N^2-1} & \frac{N (U-V)}{4}  & -\frac{N (2 T+U+V)}{4}  & 0 & \frac{N (N+3) (U-V)}{4 (N+1)} & \frac{(N-3) N (U-V)}{4(N-1)} \\
 0 & \frac{N (U-V)}{N^2-4} & 0 & -\frac{N (U+V) }{2} & \frac{N (N+3) (U-V)}{4 (N+2)} & \frac{(N-3) N (U-V)}{4 (N-2)} \\
 0 & 0 & \frac{U-V}{N} & \frac{\left(N^2-N-2\right) (U-V)}{2 N} & \frac{(2 T-(N+1) (U+V))}{2}  & 0 \\
 0 & 0 & \frac{U-V}{N} & \frac{\left(N^2+N-2\right) (U-V)}{2 N} & 0 & \frac{(-2 T-(N-1) (U+V))}{2} 
\end{array}
\right)\nonumber
\label{eq:gggg5}
\end{eqnarray}
where again 
$T=\Omega_{ab}+\Omega_{cd}$,
$U=\Omega_{ad}+\Omega_{bc}$ and 
$V=\Omega_{ac}+\Omega_{bd}$.
The $\Omega_{ij}$ factors are as in \eqref{eq:PhaseSpace}, but with an
extra minus sign for each outgoing gluon from the definition of the
triple gluon vertex, which here, and throughout this paper, is taken to be
\begin{eqnarray}
& &f_{eig} \,\,\,\,\,\mbox{    with} \nonumber \\
& &e=\mbox{the external (incoming or outgoing) eikonal gluon index}\nonumber \\
& &i=\mbox{the internal (incoming or outgoing) eikonal gluon index} \\
\label{eq:gggSign}
& &g=\mbox{the soft exchange gluon index.}\nonumber
\end{eqnarray}
The advantage of using this convention is that the color algebra does not 
depend on how the diagram is drawn on a paper, 
and is also independent of whether a gluon is incoming or outgoing.
The price to pay is an overall minus sign for each outgoing gluon compared
to \eqref{eq:PhaseSpace}.
Tensors 7 to 9 form another decoupled diagonal block
\begin{eqnarray}
\G^{3 \times 3}_{gg \to gg}=-\frac{N}{2} \times \mbox{Diagonal}[T+U,T+V,U+V].
\label{eq:3x3}
\end{eqnarray}
One way of understanding the decoupling of the $6\times 6$-block is to note
that the rules, in terms of $f_{abc}$ and $d_{abc}$, that are used when 
calculating the effect of gluon exchange, have the property that they change
the number of $f$s and $d$s only in units of two. The last three tensors 
in \eqref{eq:ggBasis}, which have one $d$ and one $f$ will therefore decouple.

The above matrices would have been symmetric in an orthonormal basis. 
As the basis \eqref{eq:ggBasis} is not normalized, when calculating the 
cross section, the scalar product of the color vectors has to be used, 
$\sigma= \M^\dagger \Sv_{gg \to gg} \M$ 
with

\begin{eqnarray}
\Sv_{gg \to gg}&=&\mbox{Diagonal}
[ 
1,
N^2-1,N^2-1,
\frac{(N^2-1)(N^2-4)}{2},
\frac{N^2 (N-1)(N+3)}{4},
\nonumber \\
& & \frac{N^2(N+1)(N-3)}{4},
2(N^2-4)(N^2-1),
2(N^2-4)(N^2-1),
\frac{(N^2-4)(N^2-1)}{2}
].\nonumber \\
\label{eq:Sgg}
\end{eqnarray}

\section{2 $\to$ 3 processes}
\label{sec:2To3}

For $2 \rightarrow n$ scattering the method of mapping initial 
$s$-channel multiplets to final multiplets persists. However, the construction 
of the final multiplets is now more complicated, as more partons are to be 
combined.
For the purpose of investigating the color structure of $2 \to 3$
processes it is enough to consider

a) $q \qbar \to q \qbar g$

b) $gg \to q \qbar g$

c) $gg \to ggg$.\\
The other cases can be obtained by changing incoming quarks to 
outgoing anti-quarks, vice versa, and by exchanging incoming gluons for
outgoing gluons.

\subsection{Construction of complete orthogonal bases}

Color conservation implies that if the incoming partons are in a certain color 
multiplet, then so must the outgoing partons be.
As for the construction of possible multiplets for the incoming partons,
this is not more complicated than in the $2 \to 2$ case. An initial
$q \qbar$ pair could form a singlet or an octet, whereas 2 initial gluons 
may be in any of the states $1$, $8^a$, $8^s$, $10$, $\overline{10}$, $27$, 
or, for $N\geq4$, $0$.

The construction of final states is, however, more complicated
as there are in general several ways of forming, for example an octet, out
of the final partons. 
One way of systemizing the construction of the final state
multiplets is to first consider the multiplet formed by two of the three 
outgoing partons. The multiplets
$M^1$ and $M^2$ can together form a number of multiplets. Each of these,
say $M^{12}$, can be combined with the remaining multiplet 
$M^3$ to form an overall multiplet $M^{tot}$.
Clearly the result $M^{tot}$ has to match the multiplet of the incoming side.
This general strategy has been applied to the above cases and will be
detailed below.
Note that this method of multiplet combination can be generalized to 
$2 \to n$ processes by continuing the subgrouping of partons.

\subsection{$q \qbar \to q \qbar g$}
\label{sec:qqbarToqqbargCon}
In the case of $q_a \qbar_b \to q_c \qbar_d g_e$, the initial $q \qbar$ 
pair is either in a singlet or in an octet state.
For an initial singlet, the final state must also be in a total singlet, 
meaning that the final $q \qbar$ must form an octet to cancel the octet from 
the outgoing gluon, giving a color tensor proportional to 
$\delta_{ab}t^e_{cd}$.

If, on the other hand, the initial state is an octet, this can be matched by
octets constructed in three different ways in the final state. 
The final $q \qbar$ can form a singlet. In this case the initial octet
is matched by the outgoing gluon, $\delta_{cd}t^e_{ba}$. 
If the final $q \qbar$ is instead in an 
octet, this octet combined with the gluon octet, can form two different octets,
$8^a$ and $8^s$, which can match the incoming octet. For more 
than three colors there are no further states to be considered 
(clearly the gluon is then no longer an octet, but a multiplet of dimension 
$N^2-1$). There are thus four color states in the basis. Guided by
the multiplet decomposition, they may be written down as

\begin{eqnarray}
T^1_{abcde}&=& \delta_{ab}t^e_{cd} \nonumber \\
T^2_{abcde}&=& \delta_{cd}t^e_{ba} \nonumber \\
T^3_{abcde}&=& t^{m}_{ba}t^{n}_{cd} i f_{m n e} \nonumber \\
T^4_{abcde}&=& t^{m}_{ba}t^{n}_{cd} d_{m n e}.
\label{eq:qqbarqqbargBasis}
\end{eqnarray}
The normalization matrix can be found in appendix \ref{BasisTensors1}, 
\eqref{eq:Sqqbarqqbarg}.

\subsection{$gg \to q \qbar g$}
\label{sec:ggToqqbargCon}
Similarly one can construct the basis for $g_a g_b \to q_c \qbar_d g_e$.
In this case the initial two gluons can form any of the multiplets
$1$, $8^a$, $8^s$, $10$, $\overline{10}$, $27$, $0$ and this must be matched on 
the outgoing side. This can be done as:

\begin{eqnarray}
& &(ab)^1 ((cd)^8e)^1 \nonumber \\
& &(ab)^{8a} ((cd)^1 e)^{8}\nonumber \\
& &(ab)^{8s} ((cd)^1 e)^{8}\nonumber \\
& &(ab)^{8a} ((cd)^8 e)^{8a}\nonumber \\
& &(ab)^{8s} ((cd)^8 e)^{8a}\nonumber \\
& &(ab)^{8a} ((cd)^8 e)^{8s}\nonumber \\
& &(ab)^{8s} ((cd)^8 e)^{8s}\nonumber \\
& &(ab)^{10} ((cd)^8 e)^{10}\nonumber \\
& &(ab)^{\overline{10}} ((cd)^8 e)^{\overline{10}}\nonumber \\
& &(ab)^{27} ((cd)^8 e)^{27}\nonumber \\
& &(ab)^{0} ((cd)^8 e)^{0}.
\label{eq:ggqqbargDecomp}
\end{eqnarray}
From this it may be concluded that there are 11 independent color tensors.
If the outgoing $q \qbar$ pair would be in an octet, there would be
9 color tensors as for $gg \to gg$, but since the $q \qbar$ pair can also
form a singlet there are two extra states, corresponding to the second 
and third case above. 
The explicit tensors are, together with the normalizations, 
given in appendix \ref{BasisTensors2}, \eqref{eq:ggqqbargBasis} and 
\eqref{eq:Sggqqbarg}.

\subsection{$gg \to ggg$}
\label{sec:ggTogggCon}
By the same principle, the basis for $gg \to ggg$ can be constructed.
This is a lengthy exercise as there turns out to be 44 independent 
color tensors.
Fortunately, due to the reasons explained in section 
\ref{sec:ggTogg}, the anomalous dimension matrix factorizes into a 22-block,
containing an odd number of $f$s (and an even number of $d$s), and 
another 22-block, containing an even number of $f$s 
(and an odd number of $d$s).

While combining the multiplet of $cd$ with the octet of $e$, 
the decompositions of 
$10 \otimes 8$,
$\overline{10} \otimes 8$,
$27 \otimes 8$ and 
$0 \otimes 8$
are needed. These may be obtained using Young tableaus.
A complication occurs for the ``decuplets'' when $N\geq4$. Apart from the 
extra 0-plets which can be combined to form tensors, there is also an
extra ``decuplet'' in $8 \otimes 10 $, and an extra ``anti-decuplet''
in $8 \otimes \overline{10}$.
Like for $gg \to gg$ the non-symmetric decuplets $10$ and $\overline{10}$
are combined as $P^{10+\overline{10}}$ and 
$P^{10-\overline{10}}$.

To explicitly see that the color tensors, written down  in 
\eqref{eq:gggggBasis}, are orthogonal, 
consider for example $T^{17}_{abcde}$. This tensor is orthogonal to 
the tensors $T^1-T^{16}$ and $T^{20}-T^{22}$ since in $T^{17}$ the gluons
$ab$ are projected onto the 27-plet whereas in 
$T^1-T^{16}$ and $T^{20}-T^{22}$ $ab$ are projected into a different multiplet.
$T^{17}$ is also orthogonal to $T^{18}$ and $T^{19}$ due to the different $cd$
projectors. Similarly one can argue about the orthogonality of all of the
tensors $T^1$ and $T^{17}-T^{22}$.
To see that all the octets ($T^{2}-T^{10}$) are mutually orthogonal one may
use the relation $f_{ABm} d_{ABn}=0$, following from the different symmetries 
w.r.t. $ab$. 
For the decuplet tensors $T^{11}-T^{16}$, 
the orthogonality of the first two and the last four follows from 
the projector argument, whereas tensor $14$ is manually constructed to be 
orthogonal to tensor $13$, which is the only decuplet for $N=3$.
Similarly the bases \eqref{eq:qqbarqqbargBasis} and \eqref{eq:ggqqbargBasis}
can be seen to be orthogonal.

\section{Results}
\label{sec:Results}

The soft anomalous dimension matrices have been calculated for the
cases $q \qbar \to q \qbar g$, $q \qbar \to ggg$ and $gg \to ggg$
using the aforementioned $s$-channel bases. 
However, the 
results are applicable to any colored five parton process by a simple 
exchange of incoming and outgoing particles. 

The sign convention for the triple gluon vertex $f_{abc}$ is calculated 
using the index order from \eqref{eq:gggSign} everywhere, implying an 
extra overall minus sign for each outgoing gluon,
compared to \eqref{eq:PhaseSpace}. 
(For a topology in which radiation is forbidden within a central rapidity 
region $Y$ and all outgoing partons have rapidity $>Y$, the momentum integrals 
of \cite{Kyrieleis:2005dt} can be used provided care is taken to compensate
for different sign conventions.)

The results presented here are calculated using a Mathematica document 
which is electronically attached.
It has been checked that the 
resulting matrices are symmetric in normalized bases 
\cite{Seymour:2005ze,Seymour:2008xr}.
It has also been explicitly checked that the bases used are complete
by projecting the color tensor arising after each possible exchange into
the basis and checking the the norm of the projected tensor is the same as 
the norm of the original tensor.
Furthermore the same piece of code has been checked to, modulo sign conventions, 
reproduce the results for several of the $2 \to 2$ cases and for 
$q q \to qqg$ in \cite{Kyrieleis:2005dt}.

The results are explicitly stated in appendix \ref{sec:AlgebraResults}, but a few
general remarks are in place. First of all, note that any gluon
exchange between the partons $ab$ must result in the same multiplet for $ab$, 
and also does not affect the $cde$ indices. Thus, an exchange between $ab$
will result in a diagonal matrix in all bases under consideration here.
Similarly a gluon exchange between $bc$ does not change the tensor.
These gluon exchanges therefore represent diagonal contributions to the soft
anomalous dimension matrices.

An exchange between the partons $ce$ or $de$, does not change the overall 
multiplet (as the initial state $ab$ is unaffected), but it may, for 
example, change a final state octet to a different final state octet. 
These contributions to $\G$ are thus block diagonal.
There is, in principle, a similar symmetry w.r.t. gluon exchange between 
$ae$ or $be$ as this does not change the multiplet of $cd$.
However, due to the ordering of basis vectors, this does not
manifest itself in a block diagonal structure.
That the above simplifications are present in the computer algebra results
offers yet another consistency check.

\section{Conclusions}
\label{sec:Conclusions}

The color structure needed for resummation of all colored five parton
processes has been calculated.
The result is a $4 \times 4$-matrix for processes involving four 
external quarks or anti-quarks and one gluon, 
an $11\times 11$-matrix for processes involving two external quarks 
or anti-quarks and three gluons, and a $22\times22$-matrix for
five gluon processes.
 
The method presented here can, by using higher projection operators,
easily be extended to processes with even 
more final state colored particles.
However, the calculations, although performed by a computer,
are quite demanding. To accomplish significant further
progress additional theoretical insight is probably needed.

\section*{Acknowledgments}
I have enjoyed discussions with Jeff Forshaw, Johan Grönqvist, 
Gösta Gustafson, Mike Seymour and Bo Söderberg. In particular the progress
was boosted by group theory discussions with Gösta Gustafson and Bo Söderberg. 
I am also extra thankful to Jeff Forshaw for introducing me to the subject and
reading my manuscript.

\bibliographystyle{utcaps}  
\bibliography{RRefs} 

\appendix

\pagebreak
\section{Basis tensors and normalizations}  
\label{BasisTensors}

\subsection{$ q \qbar \to q \qbar g $}
\label{BasisTensors1}

The basis tensors for $q_a \qbar_b \to q_c \qbar_d g_e$ are stated in \eqref{eq:qqbarqqbargBasis}. The normalizations are 

\begin{eqnarray}
  \Sv_{q \qbar \to \ q \qbar g}=
  \mbox{Diagonal}\left[
    \frac{1}{2} N \left(N^2-1\right),
    \frac{1}{2} N \left(N^2-1\right),
    \frac{1}{4} N \left(N^2-1\right),
    \frac{(N^2-4)(N^2-1)}{4N}
    \right].\nonumber\\
\label{eq:Sqqbarqqbarg}
\end{eqnarray}

\subsection{$gg \to q \qbar g$}
\label{BasisTensors2}

To explicitly write down the tensors in  \eqref{eq:ggqqbargDecomp},
the projectors in \eqref{eq:ggBasis} can be used
together with a generator component $t^g_{cd}$ to denote the 
$q \qbar$ pair in an octet. The basis tensors for 
$ g_a g_b \to q_c {\qbar}_d g_e$
may thus be chosen as:

\begin{eqnarray}
T^1_{abcde}&=&  t^e {}_{cd} \delta _{ ab }\nonumber\\
T^2_{abcde}&=& i f_{ abe } \delta _{ cd }\nonumber\\
T^3_{abcde}&=& d_{ abe } \delta _{cd}\nonumber\\
T^4_{abcde}&=& if_{ abn } if_{ me n }
    t^{m} {}_{cd}\nonumber\\
T^5_{abcde}&=& d_{ ab n } if_{ m e n }
    t^{m} {}_{cd}\nonumber\\
T^6_{abcde}&=& if_{ ab n } d_{ m e n } 
    t^{m} {}_{cd}\nonumber\\
T^7_{abcde}&=& d_{ a b n } d_{ m e n }
    t^{m} {}_{cd}\nonumber\\
T^8_{abcde}&=& P^{10+\overline{10}}_{abme}  t^{m} {}_{cd}\nonumber\\
T^9_{abcde}&=& P^{10-\overline{10}}_{abme} t^{m} {}_{cd}\nonumber\\
T^{10}_{abcde}&=&-P^{27}_{abme} t^{m} {}_{cd}\nonumber\\
T^{11}_{abcde}&=& P^{0}_{abme}t^{m} {}_{cd},
\label{eq:ggqqbargBasis}
\end{eqnarray}
where the symmetrized and anti-symmetrized decuplet projectors 
$P^{10+\overline{10}}$ and $P^{10-\overline{10}}$
are used instead of $P^{10}$ and $P^{\overline{10}}$.
The scalar products of these basis tensors are given by:
\begin{eqnarray}
& &\Sv_{gg \to q \qbar g}= \mbox{Diagonal}[\nonumber \\
& &
\frac{
\left(N^2-1\right)^2}{2},
N^2 \left(N^2-1\right),
(N^2-1)(N^2-4),
\frac{N^2 \left(N^2-1\right)}{2},
\frac{\left(N^2-4\right)\left(N^2-1\right)}{2} ,
\frac{\left(N^2-4\right) \left(N^2-1\right)}{2},
\nonumber \\
& &\frac{\left(N^2-4\right)^2 \left(N^2-1\right)}{2 N^2},
\frac{(N^2-4)(N^2-1)}{4} ,
\frac{\left(N^2-4\right) \left(N^2-1\right)}{4} ,
\frac{N^2 (N-1)(N+3)}{8} ,
\frac{N^2 (N+1)(N-3)}{8} ]\nonumber \\
\label{eq:Sggqqbarg}
\end{eqnarray}

\subsection{$gg \to ggg$}
\label{BasisTensors3}

For the $g_a g_b \to g_c g_d g_e$ 
case the physically relevant tensors are chosen as:

\begin{eqnarray}
T^1_{abcde}&=& 
i f_{c d e} \delta _{a b} \nonumber \\
T^2_{abcde}&=& 
i f_{a b e} \delta _{c d} \nonumber \\
T^3_{abcde}&=& 
-i f_{a b l} f_{c d m} f_{e l m} \nonumber \\
T^4_{abcde}&=& 
i d_{c d m} d_{e l m} f_{a b l} \nonumber \\
T^5_{abcde}&=& 
i d_{a b l} d_{e l m} f_{c d m} \nonumber \\
T^6_{abcde}&=& 
i d_{a b l} d_{c d m} f_{e l m} \nonumber \\
T^7_{abcde}&=& 
i f_{a b n}  P^{10+\overline{10}} {}_{ cden } \nonumber \\
T^8_{abcde}&=& 
d_{a b n}  P^{10-\overline{10}} {}_{ cden } \nonumber \\
T^9_{abcde}&=&
i f_{a n b}  P^{27} {}_{ cden } \nonumber \\
T^{10}_{abcde}&=& 
i f_{a b n}  P^0 {}_{ cden } \nonumber \\
T^{11}_{abcde}&=& 
i f_{c d n}  P^{10+\overline{10}} {}_{ neab } \nonumber \\
T^{12}_{abcde}&=& 
d_{c d n}  P^{10-\overline{10}} {}_{ neab } \nonumber \\
T^{13}_{abcde}&=& 
i f_{l e n}  P^{10+\overline{10}} {}_{ abmn }  P^{10+\overline{10}} {}_{ cdlm } \nonumber \\
T^{14}_{abcde}&=& d_{l e n}  P^{10-\overline{10}} {}_{ abmn }  P^{10+\overline{10}} {}_{ cdlm } \nonumber \\
&+&\frac{1}{N} i f_{l e n}  P^{10+\overline{10}} {}_{ abmn }  P^{10+\overline{10}} {}_{ cdlm } \nonumber \\
T^{15}_{abcde}&=& 
i f_{e n l }  P^{27} {}_{ cdlm }  P^{10+\overline{10}} {}_{ abmn } \nonumber \\
T^{16}_{abcde}&=& 
i f_{e l n}  P^{10+\overline{10}} {}_{ abmn }  P^0 {}_{ cdlm } \nonumber \\
T^{17}_{abcde}&=& 
i f_{c n d}  P^{27} {}_{ neab } \nonumber \\
T^{18}_{abcde}&=& 
i f_{e n l}  P^{27} {}_{ abmn }  P^{10+\overline{10}} {}_{ cdlm } \nonumber \\
T^{19}_{abcde}&=& 
i f_{e l n}  P^{27} {}_{ abmn }  P^{27} {}_{ cdlm } \nonumber \\
T^{20}_{abcde}&=& 
i f_{c d n}  P^0 {}_{ abne }  \nonumber \\
T^{21}_{abcde}&=& 
i f_{e l n}  P^{10+\overline{10}} {}_{ cdlm }  P^0 {}_{ abmn } \nonumber \\
T^{22}_{abcde}&=& 
i f_{e l n}  P^0 {}_{ abmn }  P^0 {}_{ cdlm } 
\label{eq:gggggBasis}
\end{eqnarray}
There is also a decoupled block with equally many basis tensors. 
The tensors in this block can be obtained by
$if \leftrightarrow d$ in all tensors above except $T^{14}$ in which the 
coefficient $1/N$ in front of the second term also has to be replaced by
$N/(N^2-8)$.

The normalization matrix is:

\begin{eqnarray}
& &\Sv_{gg \to ggg}= \mbox{Diagonal} \nonumber \\
& &
\left[ \right.
N \left(N^2-1\right)^2,
N \left(N^2-1\right)^2,
N^3\left(N^2-1\right),
\frac{\left(N^2-4\right)^2\left(N^2-1\right)}{N},
\frac{\left(N^2-4\right)^2\left(N^2-1\right)}{N},\nonumber \\
& &\frac{\left(N^2-4\right)^2\left(N^2-1\right)}{N},
\frac{1}{2} N (N^2-1)(N^2-4),
\frac{\left(N^2-4\right)^2 \left(N^2-1\right)}{2N},
\frac{1}{4} N^3 (N-1)(N+3),\nonumber \\
& &\frac{1}{4} N^3 (N+1)(N-3),
\frac{1}{2} N (N^2-1)(N^2-4),
\frac{\left(N^2-4\right)^2 \left(N^2-1\right)}{2N},
\frac{1}{4} N (N^2-1)(N^2-4),\nonumber\\
& &\frac{(N^2-9)(N^2-4)(N^2-1)}{4 N},
\frac{1}{8} N (N^2-1)(N+3)(N-2),
\frac{1}{8} N (N^2-1)(N+2)(N-3),\nonumber \\
& &\frac{1}{4} N^3 (N+3)(N-1),
\frac{1}{8} N(N^2-1)(N+3)(N-2),
\frac{1}{8} N^2 (N^2-1)(N+3),\nonumber \\
& &\frac{1}{4} N^3 (N+1)(N-3),
\frac{1}{8} N (N^2-1)(N+2)(N-3),
\frac{1}{8} N^2 (N^2-1)(N-3)
\left. \right].
\label{eq:Sggggg}
\end{eqnarray}

\section{Algebraic Results}
\label{sec:AlgebraResults}
Below are the algebraic results for the color part of the anomalous
dimension matrix in the notation $p_1 p_2 \to p_3 p_4 p_5$
for the processes $q \qbar \to q \qbar g$, $gg \to q \qbar g$ 
and $gg \to ggg$.

\subsection{Results for $q \qbar \to q \qbar g$}
The anomalous dimension matrix, using the basis \eqref{eq:qqbarqqbargBasis} 
is calculated to be:

\begin{eqnarray}
\Gamma_{q \qbar \rightarrow q \qbar g}&=& 
\left(
\begin{array}{llll}
 \frac{N^2-1}{2 N} & 0 & 0 & 0 \\
 0 & -\frac{1}{2 N} & 0 & 0 \\
 0 & 0 & -\frac{1}{2 N} & 0 \\
 0 & 0 & 0 & -\frac{1}{2 N}
\end{array}
\right) \Omega _{12}
+
\left(
\begin{array}{llll}
 -\frac{1}{2 N} & 0 & 0 & 0 \\
 0 & \frac{N^2-1}{2 N} & 0 & 0 \\
 0 & 0 & -\frac{1}{2 N} & 0 \\
 0 & 0 & 0 & -\frac{1}{2 N}
\end{array}
\right) \Omega _{34} 
\nonumber \\
&+&
\left(
\begin{array}{llll}
 0 & \frac{1}{2 N} & -\frac{1}{4} & \frac{1}{4}-\frac{1}{N^2} \\
 \frac{1}{2 N} & 0 & -\frac{1}{4} & \frac{1}{4}-\frac{1}{N^2} \\
 -\frac{1}{2} & -\frac{1}{2} & \frac{N^2-2}{4 N} & \
\frac{1}{N}-\frac{N}{4} \\
 \frac{1}{2} & \frac{1}{2} & -\frac{N}{4} & \frac{N^2-6}{4 N}
\end{array}
\right) \Omega _{13}
+
\left(
\begin{array}{llll}
 0 & \frac{1}{2 N} & \frac{1}{4} & \frac{1}{4}-\frac{1}{N^2} \\
 \frac{1}{2 N} & 0 & -\frac{1}{4} & \frac{1}{4}-\frac{1}{N^2} \\
 \frac{1}{2} & -\frac{1}{2} & -\frac{1}{2 N} & 0 \\
 \frac{1}{2} & \frac{1}{2} & 0 & -\frac{3}{2 N}
\end{array}
\right) \Omega _{14}
\nonumber \\
&+&
\left(
\begin{array}{llll}
 0 & \frac{1}{2 N} & -\frac{1}{4} & \frac{1}{4}-\frac{1}{N^2} \\
 \frac{1}{2 N} & 0 & \frac{1}{4} & \frac{1}{4}-\frac{1}{N^2} \\
 -\frac{1}{2} & \frac{1}{2} & -\frac{1}{2 N} & 0 \\
 \frac{1}{2} & \frac{1}{2} & 0 & -\frac{3}{2 N}
\end{array}
\right) \Omega _{23}
+
\left(
\begin{array}{llll}
 0 & \frac{1}{2 N} & \frac{1}{4} & \frac{1}{4}-\frac{1}{N^2} \\
 \frac{1}{2 N} & 0 & \frac{1}{4} & \frac{1}{4}-\frac{1}{N^2} \\
 \frac{1}{2} & \frac{1}{2} & \frac{N^2-2}{4 N} & \
\frac{N}{4}-\frac{1}{N} \\
 \frac{1}{2} & \frac{1}{2} & \frac{N}{4} & \frac{N^2-6}{4 N}
\end{array}
\right) \Omega _{24}
\nonumber \\
&+&
\left(
\begin{array}{llll}
 0 & 0 & -\frac{1}{2} & 0 \\
 0 & -\frac{N}{2} & 0 & 0 \\
 -1 & 0 & -\frac{N}{4} & \frac{1}{N}-\frac{N}{4} \\
 0 & 0 & -\frac{N}{4} & -\frac{N}{4}
\end{array}
\right) \Omega _{15}
+
\left(
\begin{array}{llll}
 0 & 0 & -\frac{1}{2} & 0 \\
 0 & \frac{N}{2} & 0 & 0 \\
 -1 & 0 & \frac{N}{4} & \frac{1}{N}-\frac{N}{4} \\
 0 & 0 & -\frac{N}{4} & \frac{N}{4}
\end{array}
\right) \Omega _{25}
\nonumber \\
&+&
\left(
\begin{array}{llll}
 \frac{N}{2} & 0 & 0 & 0 \\
 0 & 0 & \frac{1}{2} & 0 \\
 0 & 1 & \frac{N}{4} & \frac{N}{4}-\frac{1}{N} \\
 0 & 0 & \frac{N}{4} & \frac{N}{4}
\end{array}
\right) \Omega _{35}
+
\left(
\begin{array}{llll}
 -\frac{N}{2} & 0 & 0 & 0 \\
 0 & 0 & \frac{1}{2} & 0 \\
 0 & 1 & -\frac{N}{4} & \frac{N}{4}-\frac{1}{N} \\
 0 & 0 & \frac{N}{4} & -\frac{N}{4}
\end{array}
\right) \Omega _{45}
\end{eqnarray}
The momentum integrals $\Omega_{ij}$ are as in 
\eqref{eq:PhaseSpace}, including the overall minus sign for exchanges 
involving the anti-quark or the final state gluon.

Note the diagonal form of the exchange between parton pairs
in definite multiplets, that is the 12 and 34 components of 
$\G_{q \qbar \to q \qbar g}$. Also, note the block diagonal
form of the exchange between $15$ and $25$.

\subsection{Result for $gg \to q \bar{q} g$}  
\label{ggqqbarg}
Below the results for the soft anomalous dimension matrix
\begin{equation}
\G_{gg \to q \qbar g}=\sum_{i < j}\Omega_{ij}C^{ij}_{gg \to q \qbar g}
\end{equation}
are stated in terms of the basis \eqref{eq:ggqqbargBasis}.
The $\Omega_{ij}$ factors are given by \eqref{eq:PhaseSpace}, including
the overall minus sign for each involved anti-quark, and 
the overall minus sign for the outgoing gluon (arising due to the convention
for the triple gluon vertex, \eqref{eq:gggSign}, used in 
$C^{ij}_{gg \to q \qbar g}$).
\begin{eqnarray}
C^{12}_{gg \to q \qbar g}= 
\mbox{Diagonal}\left[-N,-\frac{N}{2},-\frac{N}{2},-\frac{N}{2},-\frac{N}{2},-\frac{\
N}{2},-\frac{N}{2},0,0,1,-1\right]
\end{eqnarray}
 
\begin{eqnarray}
& & C^{34}_{gg \to q \qbar g}= \\ 
& &\mbox{Diagonal}\left[-\frac{1}{2 N},\frac{N^2-1}{2 N},\frac{N^2-1}{2 N},
-\frac{1}{2 N},-\frac{1}{2 N},-\frac{1}{2 N},-\frac{1}{2 N},
-\frac{1}{2 N},-\frac{1}{2 N},-\frac{1}{2 N},-\frac{1}{2 N}\right]\nonumber 
\end{eqnarray}
\begin{minipage}[]{8cm}
\begin{eqnarray}
& & C^{35}_{gg \rightarrow q \qbar g}= \\ 
& &\left(
\begin{array}{lllllllllll}
 \frac{N}{2} & 0 & 0 & 0 & 0 & 0 & 0 & 0 & 0 & 0 & 0 \\
 0 & 0 & 0 & \frac{1}{2} & 0 & 0 & 0 & 0 & 0 & 0 & 0 \\
 0 & 0 & 0 & 0 & \frac{1}{2} & 0 & 0 & 0 & 0 & 0 & 0 \\
 0 & 1 & 0 & \frac{N}{4} & 0 & \frac{N}{4}-\frac{1}{N} & 0 & 0 & 0 & \
0 & 0 \\
 0 & 0 & 1 & 0 & \frac{N}{4} & 0 & \frac{N}{4}-\frac{1}{N} & 0 & 0 & \
0 & 0 \\
 0 & 0 & 0 & \frac{N}{4} & 0 & \frac{N}{4} & 0 & 0 & 0 & 0 & 0 \\
 0 & 0 & 0 & 0 & \frac{N}{4} & 0 & \frac{N}{4} & 0 & 0 & 0 & 0 \\
 0 & 0 & 0 & 0 & 0 & 0 & 0 & 0 & \frac{1}{2} & 0 & 0 \\
 0 & 0 & 0 & 0 & 0 & 0 & 0 & \frac{1}{2} & 0 & 0 & 0 \\
 0 & 0 & 0 & 0 & 0 & 0 & 0 & 0 & 0 & -\frac{1}{2} & 0 \\
 0 & 0 & 0 & 0 & 0 & 0 & 0 & 0 & 0 & 0 & \frac{1}{2}
\end{array}
\right)\nonumber
\end{eqnarray}
\end{minipage}
\begin{minipage}[]{8cm}
\begin{eqnarray}
& & C^{45}_{gg \rightarrow q \qbar g}= \\ 
& &\left(
\begin{array}{lllllllllll}
 -\frac{N}{2} & 0 & 0 & 0 & 0 & 0 & 0 & 0 & 0 & 0 & 0 \\
 0 & 0 & 0 & \frac{1}{2} & 0 & 0 & 0 & 0 & 0 & 0 & 0 \\
 0 & 0 & 0 & 0 & \frac{1}{2} & 0 & 0 & 0 & 0 & 0 & 0 \\
 0 & 1 & 0 & -\frac{N}{4} & 0 & \frac{N}{4}-\frac{1}{N} & 0 & 0 & 0 & \
0 & 0 \\
 0 & 0 & 1 & 0 & -\frac{N}{4} & 0 & \frac{N}{4}-\frac{1}{N} & 0 & 0 & \
0 & 0 \\
 0 & 0 & 0 & \frac{N}{4} & 0 & -\frac{N}{4} & 0 & 0 & 0 & 0 & 0 \\
 0 & 0 & 0 & 0 & \frac{N}{4} & 0 & -\frac{N}{4} & 0 & 0 & 0 & 0 \\
 0 & 0 & 0 & 0 & 0 & 0 & 0 & 0 & \frac{1}{2} & 0 & 0 \\
 0 & 0 & 0 & 0 & 0 & 0 & 0 & \frac{1}{2} & 0 & 0 & 0 \\
 0 & 0 & 0 & 0 & 0 & 0 & 0 & 0 & 0 & \frac{1}{2} & 0 \\
 0 & 0 & 0 & 0 & 0 & 0 & 0 & 0 & 0 & 0 & -\frac{1}{2}
\end{array}
\right)\nonumber
\end{eqnarray}
\end{minipage}
\begin{minipage}[]{14cm}
\begin{eqnarray}
& & C^{15}_{gg \rightarrow q \qbar g}= \\ 
& &\left(
\begin{array}{lllllllllll}
 0 & 0 & 0 & -\frac{N^2}{N^2-1} & 0 & 0 & 0 & 0 & 0 & 0 & 0 \\
 0 & -\frac{N}{2} & 0 & 0 & 0 & 0 & 0 & 0 & 0 & 0 & 0 \\
 0 & 0 & -\frac{N}{2} & 0 & 0 & 0 & 0 & 0 & 0 & 0 & 0 \\
 -1 & 0 & 0 & -\frac{N}{4} & 0 & 0 & \frac{1}{N}-\frac{N}{4} & 0 & 0 \
& \frac{N+3}{4 N+4} & -\frac{N-3}{4 (N-1)} \\
 0 & 0 & 0 & 0 & -\frac{N}{4} & -\frac{N}{4} & 0 & 0 & 0 & 0 & 0 \\
 0 & 0 & 0 & 0 & -\frac{N}{4} & -\frac{N}{4} & 0 & 0 & 0 & 0 & 0 \\
 0 & 0 & 0 & \frac{N^3}{16-4 N^2} & 0 & 0 & -\frac{N}{4} & \
\frac{N^2}{2 \left(N^2-4\right)} & 0 & 0 & 0 \\
 0 & 0 & 0 & 0 & 0 & 0 & 1 & -\frac{N}{2} & 0 & -\frac{N (N+3)}{4 \
(N+2)} & \frac{(N-3) N}{4 (N-2)} \\
 0 & 0 & 0 & 0 & 0 & 0 & 0 & 0 & -\frac{N}{2} & 0 & 0 \\
 0 & 0 & 0 & 1 & 0 & 0 & 0 & -\frac{N}{2}+\frac{1}{2}+\frac{1}{N} & 0 \
& \frac{1}{2} (-N-1) & 0 \\
 0 & 0 & 0 & -1 & 0 & 0 & 0 & \frac{N^2+N-2}{2 N} & 0 & 0 & \
\frac{1-N}{2}
\end{array}
\right)\nonumber
\end{eqnarray}
\end{minipage}

\hspace*{-1.7cm}
\begin{minipage}[]{20cm}
\hspace*{-1.0cm}
\begin{minipage}[]{14cm}
\begin{eqnarray}
& & C^{25}_{gg \rightarrow q \qbar g}= \\ 
& &\left(
\begin{array}{lllllllllll}
 0 & 0 & 0 & \frac{N^2}{N^2-1} & 0 & 0 & 0 & 0 & 0 & 0 & 0 \\
 0 & -\frac{N}{2} & 0 & 0 & 0 & 0 & 0 & 0 & 0 & 0 & 0 \\
 0 & 0 & -\frac{N}{2} & 0 & 0 & 0 & 0 & 0 & 0 & 0 & 0 \\
 1 & 0 & 0 & -\frac{N}{4} & 0 & 0 & \frac{N}{4}-\frac{1}{N} & 0 & 0 & \
-\frac{N+3}{4 N+4} & \frac{N-3}{4 (N-1)} \\
 0 & 0 & 0 & 0 & -\frac{N}{4} & \frac{N}{4} & 0 & 0 & 0 & 0 & 0 \\
 0 & 0 & 0 & 0 & \frac{N}{4} & -\frac{N}{4} & 0 & 0 & 0 & 0 & 0 \\
 0 & 0 & 0 & \frac{N^3}{4 \left(N^2-4\right)} & 0 & 0 & -\frac{N}{4} \
& \frac{N^2}{8-2 N^2} & 0 & 0 & 0 \\
 0 & 0 & 0 & 0 & 0 & 0 & -1 & -\frac{N}{2} & 0 & \frac{N (N+3)}{4 \
(N+2)} & -\frac{(N-3) N}{4 (N-2)} \\
 0 & 0 & 0 & 0 & 0 & 0 & 0 & 0 & -\frac{N}{2} & 0 & 0 \\
 0 & 0 & 0 & -1 & 0 & 0 & 0 & \frac{1}{2} \left(N-1-\frac{2}{N}\right) & 0 & \frac{1}{2} (-N-1) & 0 \\
 0 & 0 & 0 & 1 & 0 & 0 & 0 & -\frac{N}{2}-\frac{1}{2}+\frac{1}{N} & 0 \
& 0 & \frac{1-N}{2}
\end{array}
\right)\nonumber
\end{eqnarray}
\end{minipage}\\
\hspace*{-1.7cm}
\begin{minipage}[]{20cm}
\begin{eqnarray}
& & C^{13}_{gg \rightarrow q \qbar g}= \\ 
& &\left(
\begin{array}{lllllllllll}
 0 & \frac{N}{N^2-1} & 0 & \frac{N^2}{2-2 N^2} & 0 & \frac{N^2-4}{2 \
\left(N^2-1\right)} & 0 & 0 & 0 & 0 & 0 \\
 \frac{1}{2 N} & 0 & 0 & -\frac{1}{4} & 0 & 0 & \
\frac{1}{4}-\frac{1}{N^2} & 0 & 0 & \frac{N+3}{8 N+8} & \frac{N-3}{8 \
(N-1)} \\
 0 & 0 & 0 & 0 & -\frac{1}{4} & \frac{1}{4} & 0 & 0 & \frac{1}{4} & 0 \
& 0 \\
 -\frac{1}{2} & -\frac{1}{2} & 0 & \frac{N}{8} & \frac{N^2-4}{8 N} & \
-\frac{N^2-4}{8 N} & -\frac{N^2-4}{8 N} & 0 & 0 & \frac{N+3}{8 N+8} & \
-\frac{N-3}{8 (N-1)} \\
 0 & 0 & -\frac{1}{2} & \frac{N}{8} & \frac{N}{8} & -\frac{N}{8} & \
-\frac{N^2-4}{8 N} & \frac{1}{4} & 0 & 0 & 0 \\
 \frac{1}{2} & 0 & \frac{1}{2} & -\frac{N}{8} & -\frac{N}{8} & \
\frac{N}{8} & \frac{N^2-12}{8 N} & 0 & 0 & \frac{N (N+3)}{8 \
\left(N^2+3 N+2\right)} & -\frac{(N-3) N}{8 \left(N^2-3 N+2\right)} \
\\
 0 & \frac{N^2}{2 \left(N^2-4\right)} & 0 & \frac{N^3}{32-8 N^2} & \
-\frac{N}{8} & \frac{N \left(N^2-12\right)}{8 \left(N^2-4\right)} & \
\frac{N}{8} & \frac{N^2}{4 \left(N^2-4\right)} & \frac{N}{8-2 N^2} & \
0 & 0 \\
 0 & 0 & 0 & 0 & \frac{1}{2} & 0 & \frac{1}{2} & \frac{N}{4} & \
-\frac{1}{4} & -\frac{N (N+3)}{8 (N+2)} & \frac{(N-3) N}{8 (N-2)} \\
 0 & 0 & 1 & 0 & 0 & 0 & -\frac{1}{N} & -\frac{1}{4} & \frac{N}{4} & \
-\frac{N (N+3)}{8 (N+2)} & -\frac{(N-3) N}{8 (N-2)} \\
 0 & 1 & 0 & \frac{1}{2} & 0 & \frac{1}{2}-\frac{1}{N} & 0 & \
\frac{-N^2+N+2}{4 N} & \frac{-N^2+N+2}{4 N} & \frac{N+1}{4} & 0 \\
 0 & 1 & 0 & -\frac{1}{2} & 0 & -\frac{N+2}{2 N} & 0 & \
\frac{N^2+N-2}{4 N} & \frac{1}{4} \left(-N-1+\frac{2}{N}\right) & 0 & \
\frac{N-1}{4}
\end{array}
\right)\nonumber
\end{eqnarray}
\end{minipage}\\
\hspace*{-1.5cm}
\begin{minipage}[]{20cm}
\begin{eqnarray}
& & C^{23}_{gg \rightarrow q \qbar g}= \\
& &\left(
\begin{array}{lllllllllll}
 0 & -\frac{N}{N^2-1} & 0 & \frac{N^2}{2 \left(N^2-1\right)} & 0 & \
-\frac{N^2-4}{2 \left(N^2-1\right)} & 0 & 0 & 0 & 0 & 0 \\
 -\frac{1}{2 N} & 0 & 0 & -\frac{1}{4} & 0 & 0 & \
\frac{1}{N^2}-\frac{1}{4} & 0 & 0 & -\frac{N+3}{8 N+8} & \
-\frac{N-3}{8 (N-1)} \\
 0 & 0 & 0 & 0 & -\frac{1}{4} & -\frac{1}{4} & 0 & 0 & -\frac{1}{4} & \
0 & 0 \\
 \frac{1}{2} & -\frac{1}{2} & 0 & \frac{N}{8} & -\frac{N^2-4}{8 N} & \
-\frac{N^2-4}{8 N} & \frac{N^2-4}{8 N} & 0 & 0 & -\frac{N+3}{8 N+8} & \
\frac{N-3}{8 (N-1)} \\
 0 & 0 & -\frac{1}{2} & -\frac{N}{8} & \frac{N}{8} & \frac{N}{8} & \
-\frac{N^2-4}{8 N} & -\frac{1}{4} & 0 & 0 & 0 \\
 -\frac{1}{2} & 0 & -\frac{1}{2} & -\frac{N}{8} & \frac{N}{8} & \
\frac{N}{8} & \frac{3}{2 N}-\frac{N}{8} & 0 & 0 & -\frac{N (N+3)}{8 \
\left(N^2+3 N+2\right)} & \frac{(N-3) N}{8 \left(N^2-3 N+2\right)} \\
 0 & \frac{N^2}{8-2 N^2} & 0 & \frac{N^3}{8 \left(N^2-4\right)} & \
-\frac{N}{8} & -\frac{N \left(N^2-12\right)}{8 \left(N^2-4\right)} & \
\frac{N}{8} & \frac{N^2}{16-4 N^2} & \frac{N}{2 \left(N^2-4\right)} & \
0 & 0 \\
 0 & 0 & 0 & 0 & -\frac{1}{2} & 0 & -\frac{1}{2} & \frac{N}{4} & \
-\frac{1}{4} & \frac{N (N+3)}{8 (N+2)} & -\frac{(N-3) N}{8 (N-2)} \\
 0 & 0 & -1 & 0 & 0 & 0 & \frac{1}{N} & -\frac{1}{4} & \frac{N}{4} & \
\frac{N (N+3)}{8 (N+2)} & \frac{(N-3) N}{8 (N-2)} \\
 0 & -1 & 0 & -\frac{1}{2} & 0 & \frac{1}{N}-\frac{1}{2} & 0 & \
-\frac{-N^2+N+2}{4 N} & -\frac{-N^2+N+2}{4 N} & \frac{N+1}{4} & 0 \\
 0 & -1 & 0 & \frac{1}{2} & 0 & \frac{1}{2}+\frac{1}{N} & 0 & \
\frac{1}{4} \left(-N-1+\frac{2}{N}\right) & \frac{N^2+N-2}{4 N} & 0 & \
\frac{N-1}{4}
\end{array}
\right)\nonumber
\end{eqnarray}
\end{minipage}
\end{minipage}

\hspace*{-2.7cm}
\begin{minipage}[]{20cm}
\hspace*{-1.7cm}
\begin{minipage}[]{20cm}
\begin{eqnarray}
& & C^{14}_{gg \rightarrow q \qbar g}= \\
& &\left(
\begin{array}{lllllllllll}
 0 & \frac{N}{N^2-1} & 0 & \frac{N^2}{2 \left(N^2-1\right)} & 0 & \
\frac{N^2-4}{2 \left(N^2-1\right)} & 0 & 0 & 0 & 0 & 0 \\
 \frac{1}{2 N} & 0 & 0 & -\frac{1}{4} & 0 & 0 & \
\frac{1}{4}-\frac{1}{N^2} & 0 & 0 & \frac{N+3}{8 N+8} & \frac{N-3}{8 \
(N-1)} \\
 0 & 0 & 0 & 0 & -\frac{1}{4} & \frac{1}{4} & 0 & 0 & \frac{1}{4} & 0 \
& 0 \\
 \frac{1}{2} & -\frac{1}{2} & 0 & -\frac{N}{8} & \frac{N^2-4}{8 N} & \
-\frac{N^2-4}{8 N} & \frac{N^2-4}{8 N} & 0 & 0 & -\frac{N+3}{8 N+8} & \
\frac{N-3}{8 (N-1)} \\
 0 & 0 & -\frac{1}{2} & \frac{N}{8} & -\frac{N}{8} & \frac{N}{8} & \
-\frac{N^2-4}{8 N} & \frac{1}{4} & 0 & 0 & 0 \\
 \frac{1}{2} & 0 & \frac{1}{2} & -\frac{N}{8} & \frac{N}{8} & \
-\frac{N}{8} & \frac{N^2-12}{8 N} & 0 & 0 & \frac{N (N+3)}{8 \
\left(N^2+3 N+2\right)} & -\frac{(N-3) N}{8 \left(N^2-3 N+2\right)} \
\\
 0 & \frac{N^2}{2 \left(N^2-4\right)} & 0 & \frac{N^3}{8 \left(N^2-4\right)} & -\frac{N}{8} & \frac{N \left(N^2-12\right)}{8 \left(N^2-4\right)} & -\frac{N}{8} & \frac{N^2}{16-4 N^2} & \frac{N}{8-2 N^2} & 0 \
& 0 \\
 0 & 0 & 0 & 0 & \frac{1}{2} & 0 & -\frac{1}{2} & -\frac{N}{4} & \
-\frac{1}{4} & \frac{N (N+3)}{8 (N+2)} & -\frac{(N-3) N}{8 (N-2)} \\
 0 & 0 & 1 & 0 & 0 & 0 & -\frac{1}{N} & -\frac{1}{4} & -\frac{N}{4} & \
-\frac{N (N+3)}{8 (N+2)} & -\frac{(N-3) N}{8 (N-2)} \\
 0 & 1 & 0 & -\frac{1}{2} & 0 & \frac{1}{2}-\frac{1}{N} & 0 & \
-\frac{-N^2+N+2}{4 N} & \frac{-N^2+N+2}{4 N} & \frac{1}{4} (-N-1) & 0 \
\\
 0 & 1 & 0 & \frac{1}{2} & 0 & -\frac{N+2}{2 N} & 0 & \frac{1}{4} \
\left(-N-1+\frac{2}{N}\right) & \frac{1}{4} \left(-N-1+\frac{2}{N}\right) & 0 & \frac{1-N}{4}
\end{array}
\right) \nonumber
\end{eqnarray}
\end{minipage}
\hspace*{-0.7cm}
\begin{minipage}[]{20cm}
\begin{eqnarray}
& & C^{24}_{gg \rightarrow q \qbar g}= \\ 
& &\left(
\begin{array}{lllllllllll}
 0 & -\frac{N}{N^2-1} & 0 & \frac{N^2}{2-2 N^2} & 0 & -\frac{N^2-4}{2 \
\left(N^2-1\right)} & 0 & 0 & 0 & 0 & 0 \\
 -\frac{1}{2 N} & 0 & 0 & -\frac{1}{4} & 0 & 0 & \
\frac{1}{N^2}-\frac{1}{4} & 0 & 0 & -\frac{N+3}{8 N+8} & \
-\frac{N-3}{8 (N-1)} \\
 0 & 0 & 0 & 0 & -\frac{1}{4} & -\frac{1}{4} & 0 & 0 & -\frac{1}{4} & \
0 & 0 \\
 -\frac{1}{2} & -\frac{1}{2} & 0 & -\frac{N}{8} & -\frac{N^2-4}{8 N} \
& -\frac{N^2-4}{8 N} & -\frac{N^2-4}{8 N} & 0 & 0 & \frac{N+3}{8 N+8} \
& -\frac{N-3}{8 (N-1)} \\
 0 & 0 & -\frac{1}{2} & -\frac{N}{8} & -\frac{N}{8} & -\frac{N}{8} & \
-\frac{N^2-4}{8 N} & -\frac{1}{4} & 0 & 0 & 0 \\
 -\frac{1}{2} & 0 & -\frac{1}{2} & -\frac{N}{8} & -\frac{N}{8} & \
-\frac{N}{8} & \frac{3}{2 N}-\frac{N}{8} & 0 & 0 & -\frac{N (N+3)}{8 \
\left(N^2+3 N+2\right)} & \frac{(N-3) N}{8 \left(N^2-3 N+2\right)} \\
 0 & \frac{N^2}{8-2 N^2} & 0 & \frac{N^3}{32-8 N^2} & -\frac{N}{8} & \
-\frac{N \left(N^2-12\right)}{8 \left(N^2-4\right)} & -\frac{N}{8} & \
\frac{N^2}{4 \left(N^2-4\right)} & \frac{N}{2 \left(N^2-4\right)} & 0 \
& 0 \\
 0 & 0 & 0 & 0 & -\frac{1}{2} & 0 & \frac{1}{2} & -\frac{N}{4} & \
-\frac{1}{4} & -\frac{N (N+3)}{8 (N+2)} & \frac{(N-3) N}{8 (N-2)} \\
 0 & 0 & -1 & 0 & 0 & 0 & \frac{1}{N} & -\frac{1}{4} & -\frac{N}{4} & \
\frac{N (N+3)}{8 (N+2)} & \frac{(N-3) N}{8 (N-2)} \\
 0 & -1 & 0 & \frac{1}{2} & 0 & \frac{1}{N}-\frac{1}{2} & 0 & \
\frac{-N^2+N+2}{4 N} & -\frac{-N^2+N+2}{4 N} & \frac{1}{4} (-N-1) & 0 \
\\
 0 & -1 & 0 & -\frac{1}{2} & 0 & \frac{1}{2}+\frac{1}{N} & 0 & \
\frac{N^2+N-2}{4 N} & \frac{N^2+N-2}{4 N} & 0 & \frac{1-N}{4}
\end{array}
\right) \nonumber
\end{eqnarray}
\end{minipage}
\end{minipage}

\pagebreak

\subsection{Result for gg $\to$ ggg}
Here the color structure of gluon exchange between partons $ij$ are stated
in the basis of \eqref{eq:gggggBasis}.
The soft anomalous dimension matrix is given by

\begin{equation}
\G_{gg \to ggg}=\sum_{i <  j}\Omega_{ij}C^{ij}_{gg \to ggg},
\end{equation}
where the phase space integrals are as in \eqref{eq:PhaseSpace},
but with an extra minus sign for each outgoing gluon due to the
convention for the triple gluon vertex  from \eqref{eq:gggSign}, 
that is $\Omega_{1 \mbox{ or } 2,\mbox{ } 3\mbox{ or } 4 \mbox{ or } 5}$
has an overall minus sign compared to \eqref{eq:PhaseSpace}.
Due to their complexity, the 13, 14, 23 and 24 components are written over 
two pages.
The color matrices are given by:

\begin{eqnarray}
    & & C^{12}_{gg \to ggg}= \\
    & &\mbox{Diagonal}
    \left[ 
      -N,-\frac{N}{2},-\frac{N}{2},-\frac{N}{2},-\frac{N}{2},-\frac{N}
      {2},-\frac{N}{2},-\frac{N}{2},-\frac{N}{2},-\frac{N}{2},0,0,0,0,0,0,
      1,1,1,-1,-1,-1
      \right]\nonumber
\end{eqnarray}

\begin{eqnarray}
& &C^{34}_{gg \to ggg}= \\
& &\mbox{Diagonal}
\left[ 
-\frac{N}{2},-N,-\frac{N}{2},-\frac{N}{2},-\frac{N}{2},-\frac{N}
   {2},0,0,1,-1,-\frac{N}{2},-\frac{N}{2},0,0,1,-1,-\frac{N}{2},0,1,
   -\frac{N}{2},0,-1
\right]\nonumber
\end{eqnarray}


\pagebreak
\begin{figure}[h]
\hspace*{-2.2cm}
$C^{15}_{gg \to ggg}=$ 
\vspace*{-0.5cm}
\begin{equation}
\label{eq:GPart15}
\end{equation}
 \hspace*{-2.2cm}
    \includegraphics[width=20cm]{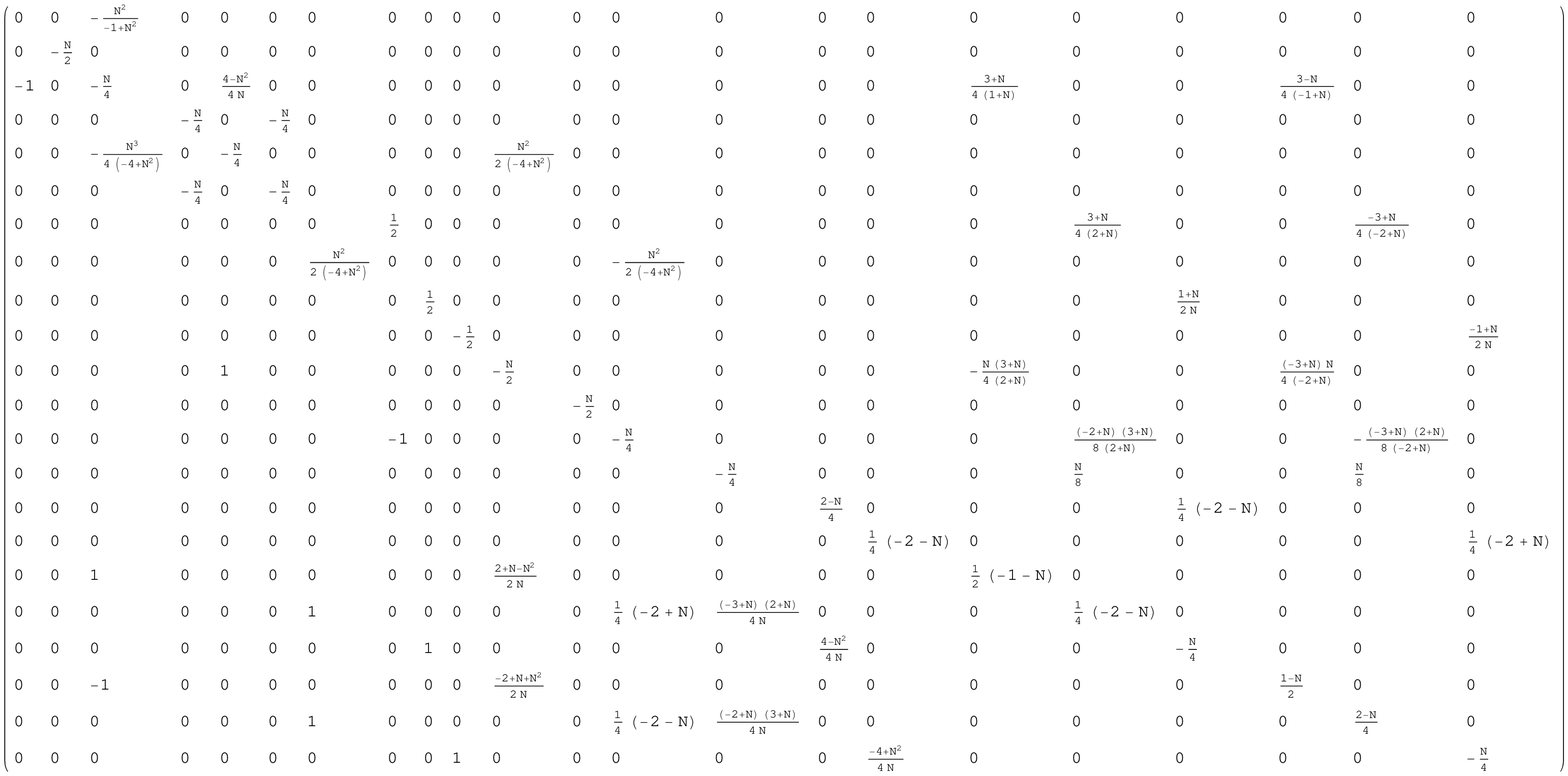}
\\
\hspace*{-2.2cm}
$C^{25}_{gg \to ggg}=$ 
\vspace*{-0.5cm}
\begin{equation}
\label{eq:GPart25}
\end{equation}
 \hspace*{-2.2cm}
    \includegraphics[width=20cm]{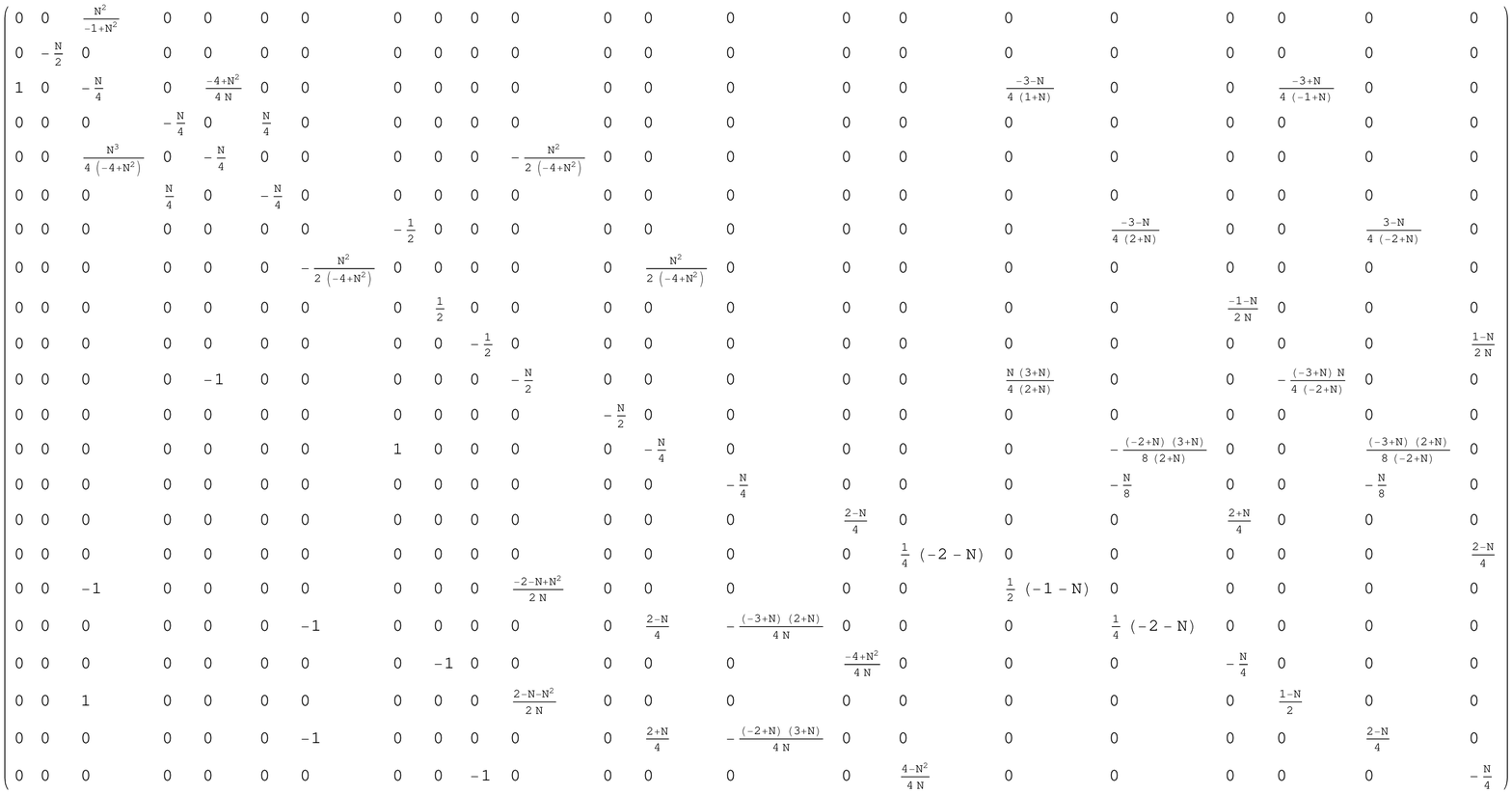}
\end{figure}

\pagebreak

\begin{figure}[h]
\hspace*{-2.2cm}
$C^{35}_{gg \to ggg}=$ 
\vspace*{-0.5cm}
\begin{equation}
\label{eq:GPart35}
\end{equation}
 \hspace*{-2.2cm}
    \includegraphics[width=20cm]{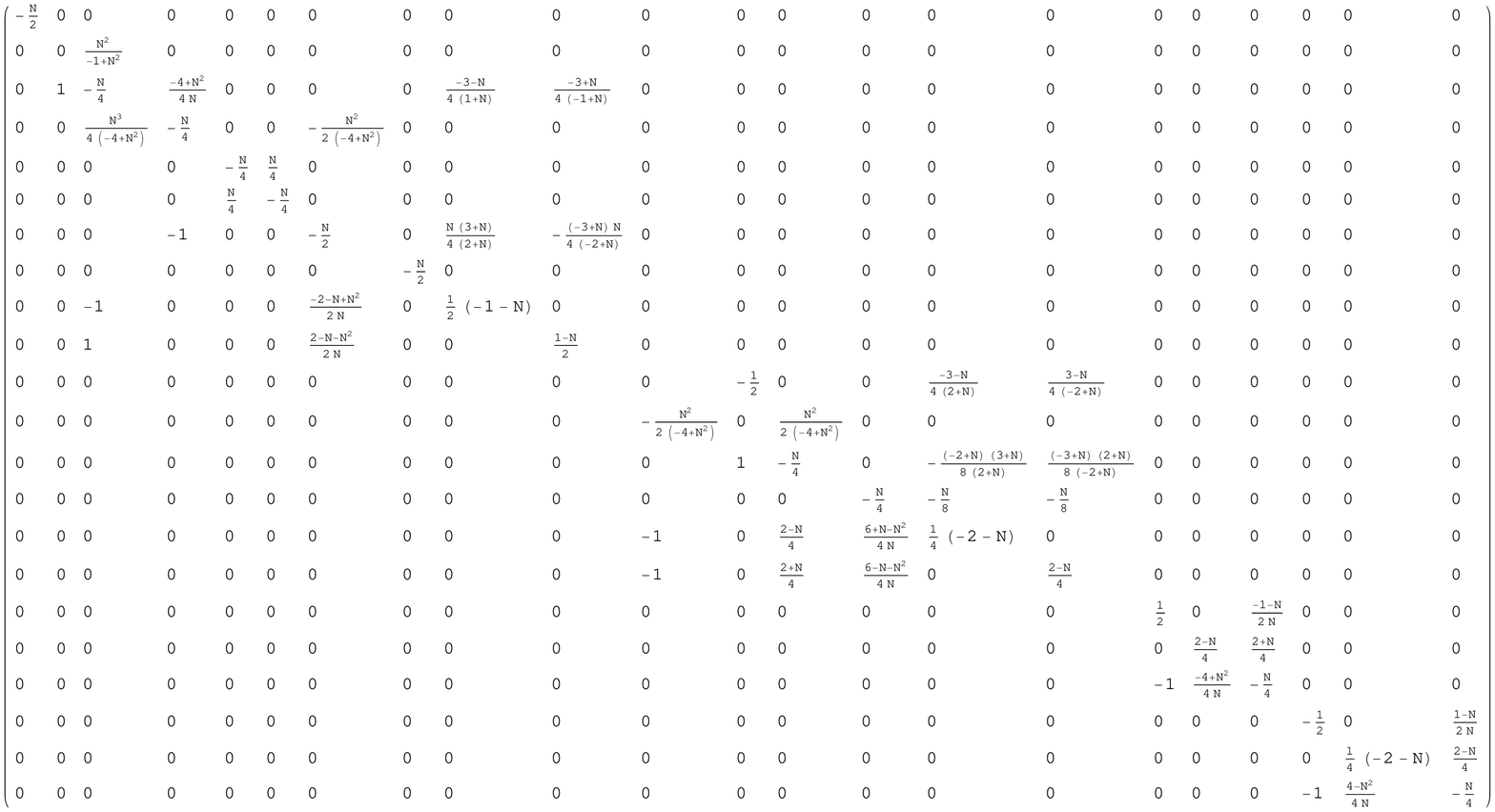}
\\
\\
\hspace*{-2.2cm}
$C^{45}_{gg \to ggg}=$ 
\vspace*{-0.5cm}
\begin{equation}
\label{eq:GPart45}
\end{equation}
 \hspace*{-2.2cm}
    \includegraphics[width=20cm]{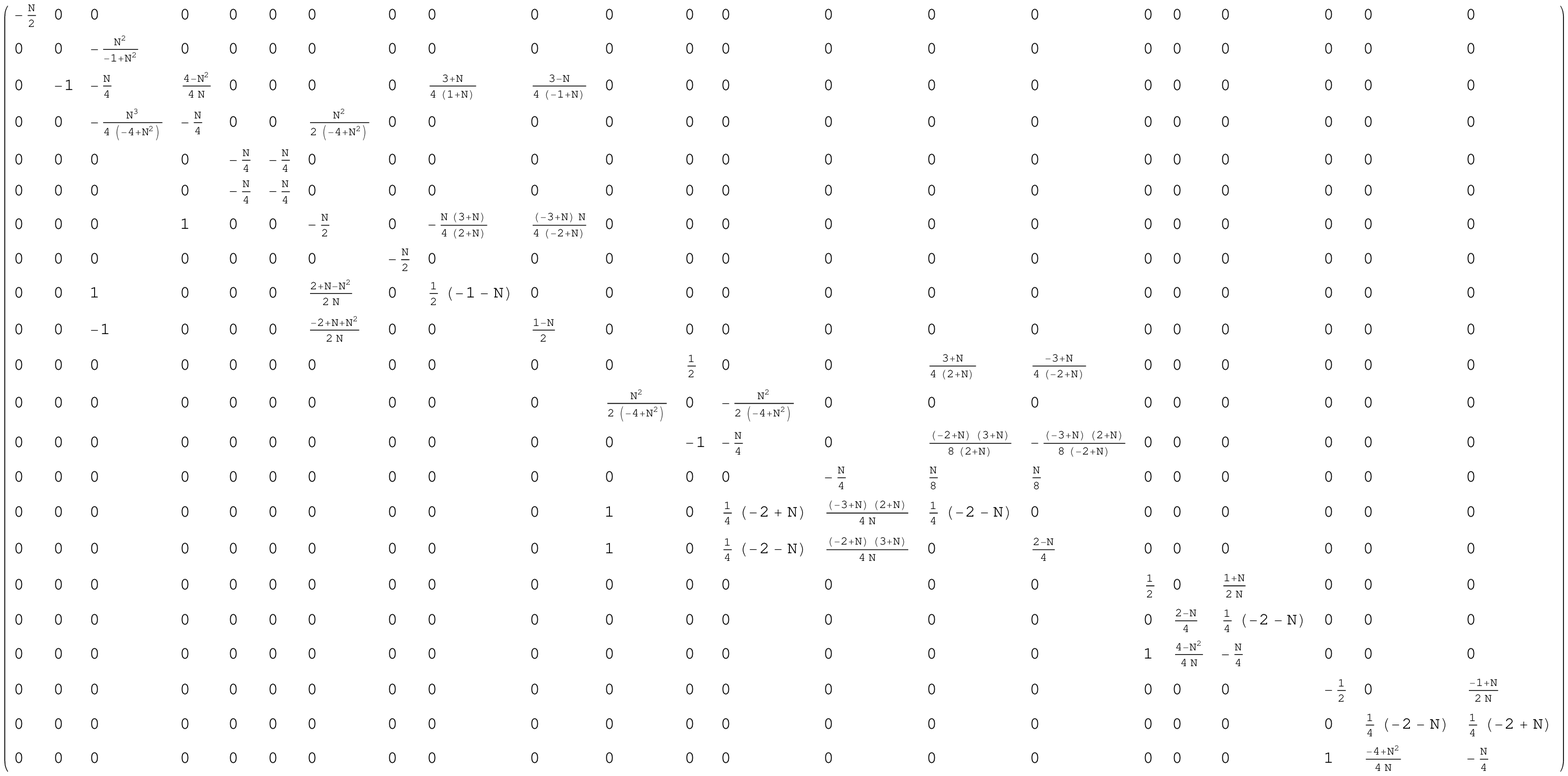}
\end{figure}



\pagebreak




\begin{figure}[]
\vspace{-0.5 cm}
$C^{13}_{gg \to ggg}=$ \\
\hspace*{-0.5 cm}
$\left( \phantom{\begin{array}{llll}
 & | & \\
 & | & \\
 & | & \\
 & | & \\
 & | & \\
 & | & \\
 & | & \\
 & | & \\
 & | & \\
 & | & \\
 & | & \\
 & | & \\
 & | & \\
 & | & \\
 & | & \\
 & | & \\
 & | & \\
 & | & \\
 & | & \\
 & | & \\
 & | & \\
 & | & \\
 & | & \\
\end{array}}\right.$
\vspace*{-12 cm}
\\
\includegraphics[width=16cm]{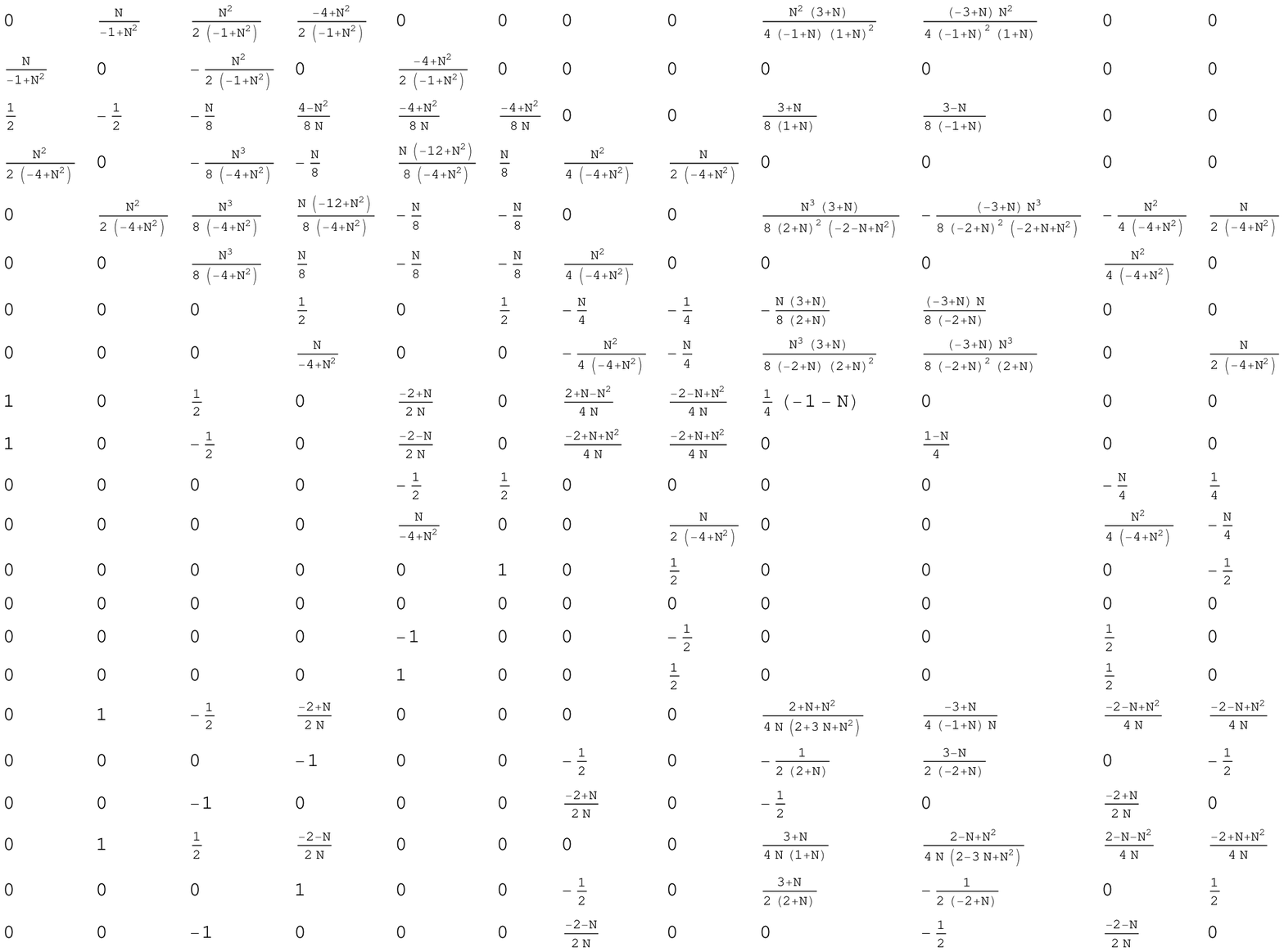}
$C^{14}_{gg \to ggg}=$ \\
\hspace*{-0.5 cm}
$\left( \phantom{\begin{array}{llll}
 & | & \\
 & | & \\
 & | & \\
 & | & \\
 & | & \\
 & | & \\
 & | & \\
 & | & \\
 & | & \\
 & | & \\
 & | & \\
 & | & \\
 & | & \\
 & | & \\
 & | & \\
 & | & \\
 & | & \\
 & | & \\
 & | & \\
 & | & \\
 & | & \\
 & | & \\
\end{array}}\right.$
\vspace*{-11.5 cm}
\\
    \includegraphics[width=16cm]{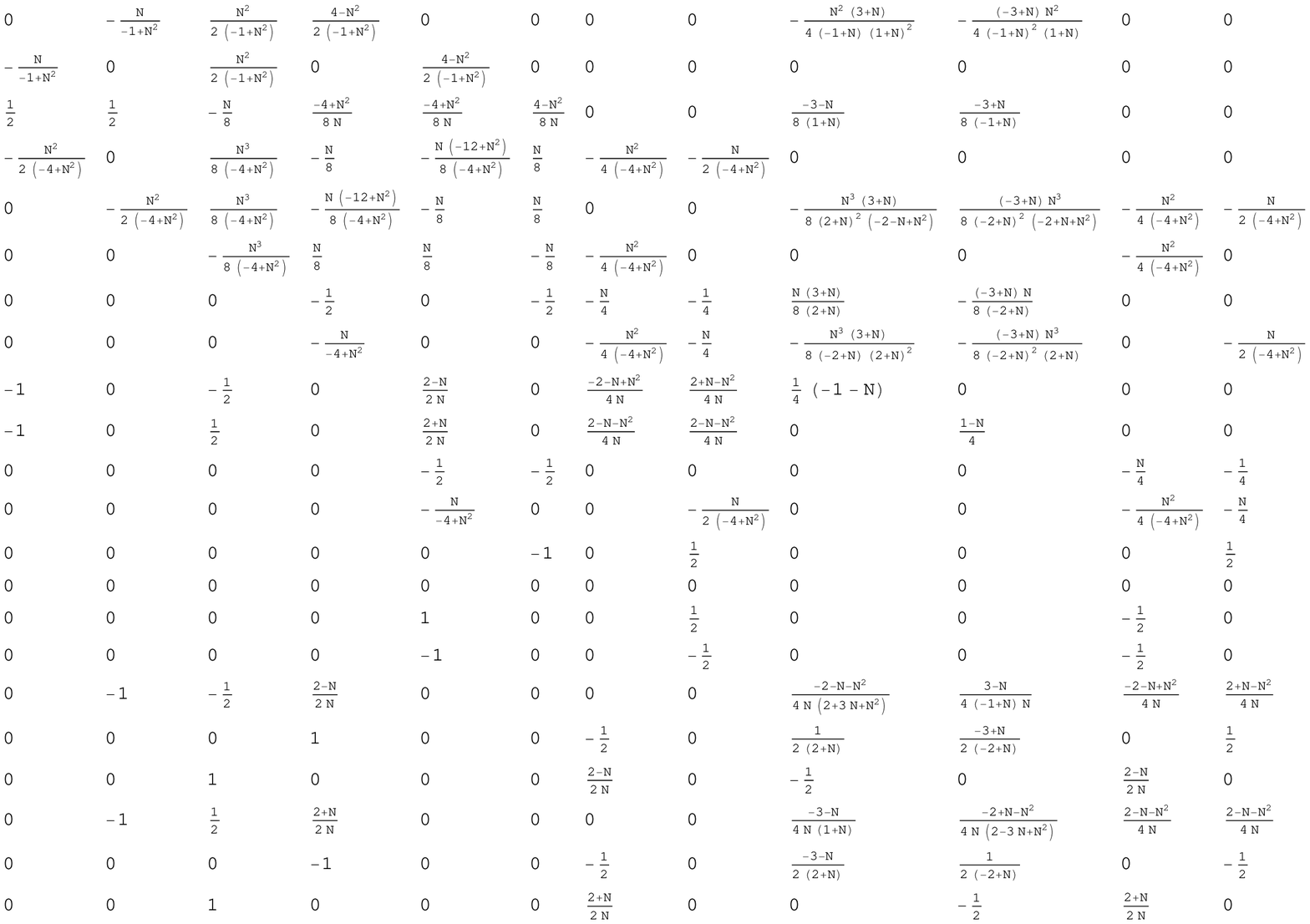}
\end{figure}

\pagebreak
\begin{figure}[h]
\begin{equation}
\label{eq:GPart13}
\end{equation}
\hspace*{15.5 cm}
$\left. \phantom{\begin{array}{llll}
 & | & \\
 & | & \\
 & | & \\
 & | & \\
 & | & \\
 & | & \\
 & | & \\
 & | & \\
 & | & \\
 & | & \\
 & | & \\
 & | & \\
 & | & \\
 & | & \\
 & | & \\
 & | & \\
 & | & \\
 & | & \\
 & | & \\
 & | & \\
 & | & \\
 & | & \\
 \end{array}}\right)$
\vspace*{-11.5 cm}
\\
    \includegraphics[width=16cm]{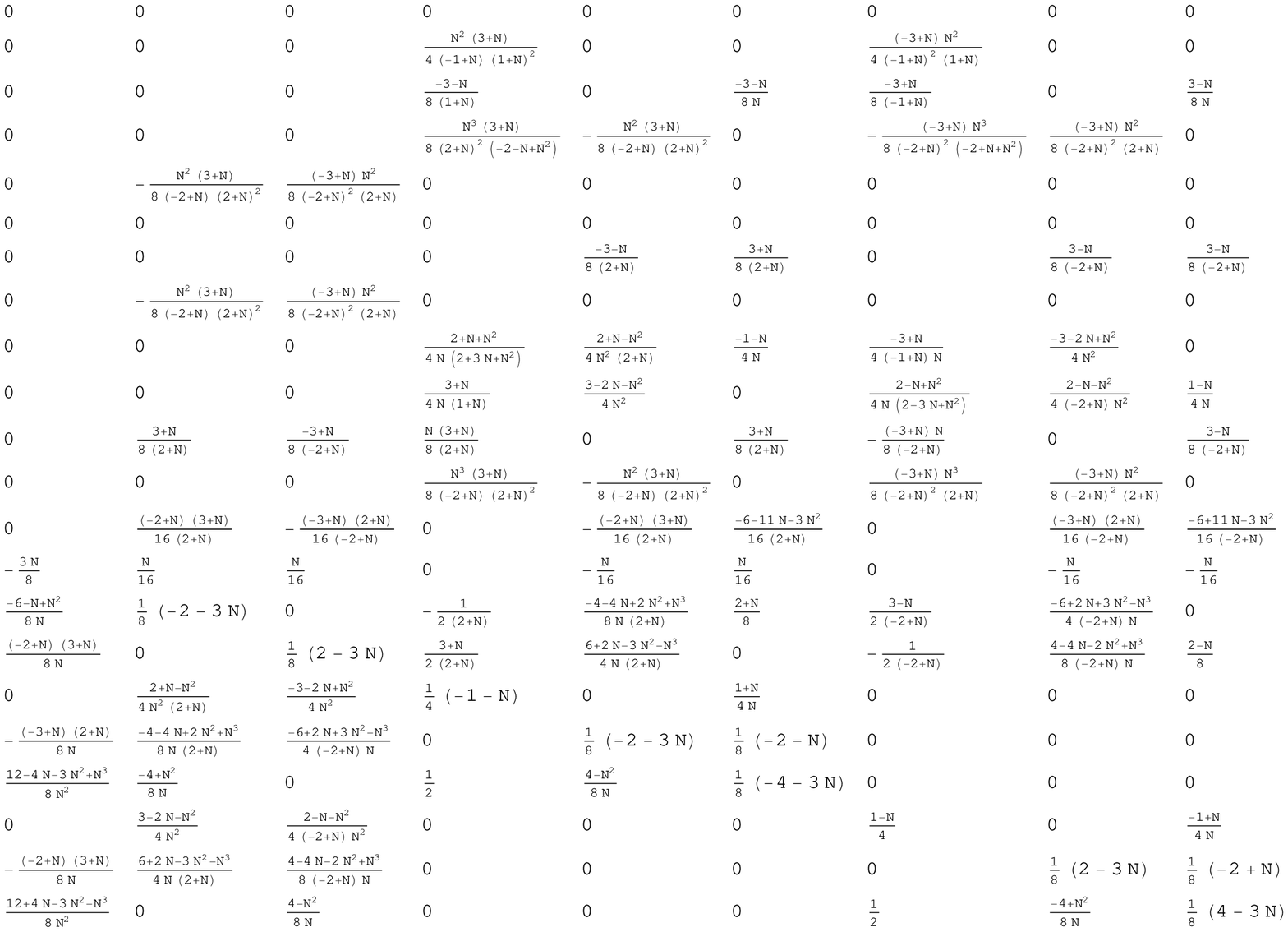}
\begin{equation}
\label{eq:GPart14}
\end{equation}
\hspace*{15.5 cm}
$\left. \phantom{\begin{array}{llll}
 & | & \\
 & | & \\
 & | & \\
 & | & \\
 & | & \\
 & | & \\
 & | & \\
 & | & \\
 & | & \\
 & | & \\
 & | & \\
 & | & \\
 & | & \\
 & | & \\
 & | & \\
 & | & \\
 & | & \\
 & | & \\
 & | & \\
 & | & \\
 & | & \\
 \end{array}}\right)$
\vspace*{-11 cm}
\\
    \includegraphics[width=16cm]{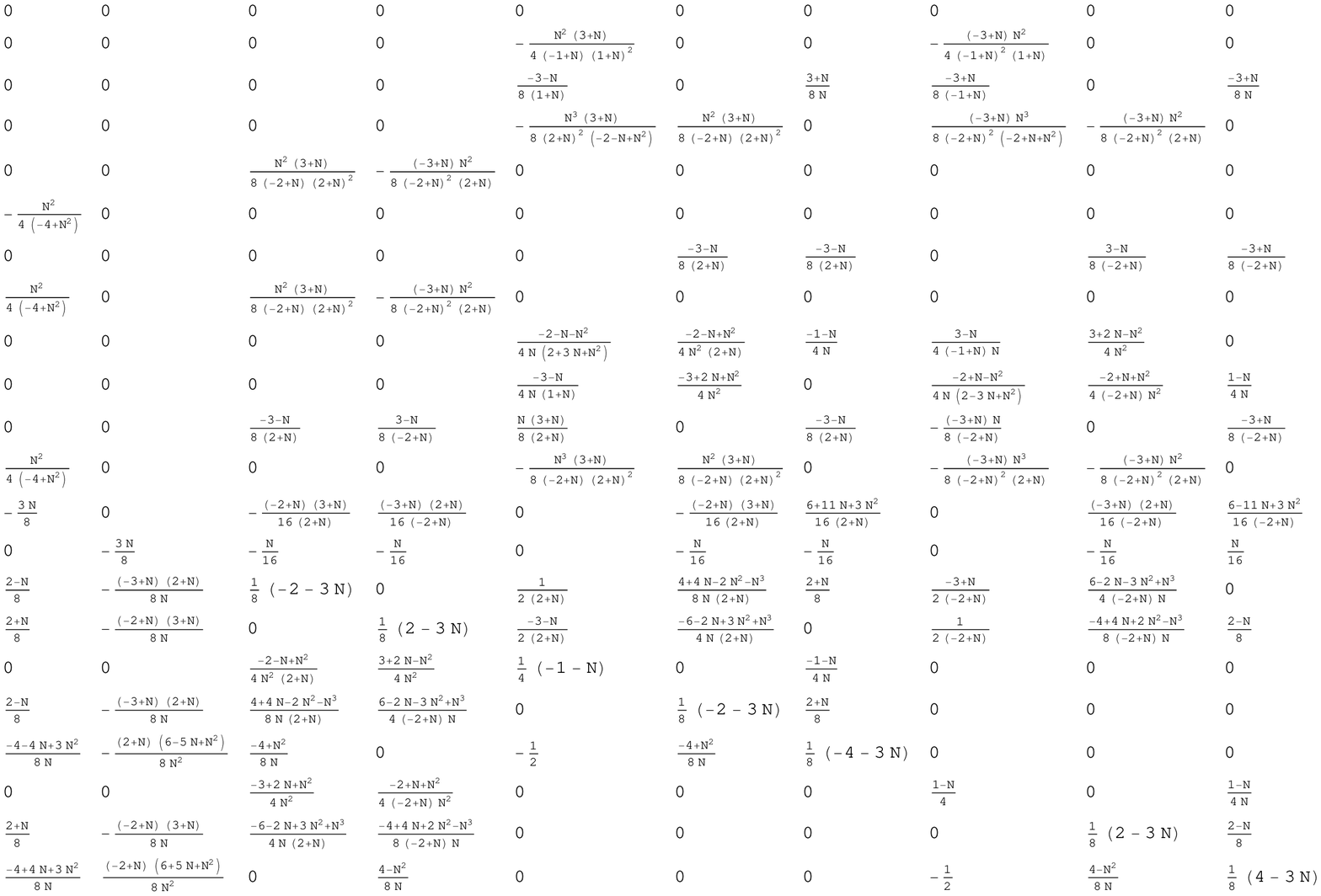}
\end{figure}

\pagebreak
\begin{figure}[]
$C^{23}_{gg \to ggg}=$ \\
\hspace*{-0.5 cm}
$\left( \phantom{\begin{array}{llll}
 & | & \\
 & | & \\
 & | & \\
 & | & \\
 & | & \\
 & | & \\
 & | & \\
 & | & \\
 & | & \\
 & | & \\
 & | & \\
 & | & \\
 & | & \\
 & | & \\
 & | & \\
 & | & \\
 & | & \\
 & | & \\
 & | & \\
 & | & \\
 \end{array}}\right.$
\vspace*{-10.5 cm}
\\
\includegraphics[width=16cm]{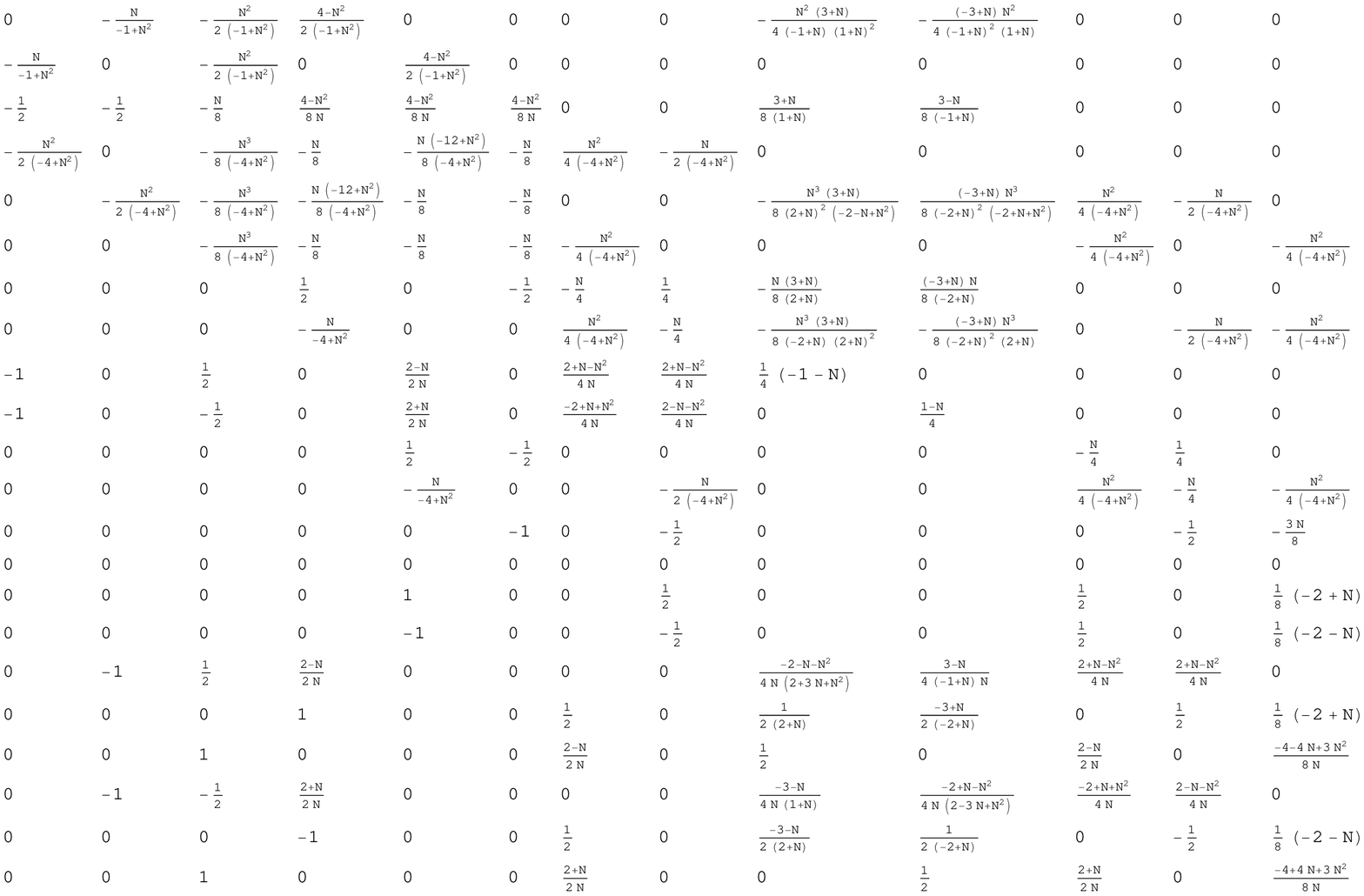}
$C^{24}_{gg \to ggg}=$ \\
\hspace*{-0.5 cm}
$\left( \phantom{\begin{array}{llll}
 & | & \\
 & | & \\
 & | & \\
 & | & \\
 & | & \\
 & | & \\
 & | & \\
 & | & \\
 & | & \\
 & | & \\
 & | & \\
 & | & \\
 & | & \\
 & | & \\
 & | & \\
 & | & \\
 & | & \\
 & | & \\
 & | & \\
 & | & \\
 & | & \\
 & | & \\
 & | & \\
\end{array}}\right.$
\vspace*{-12 cm}
\\
    \includegraphics[width=16cm]{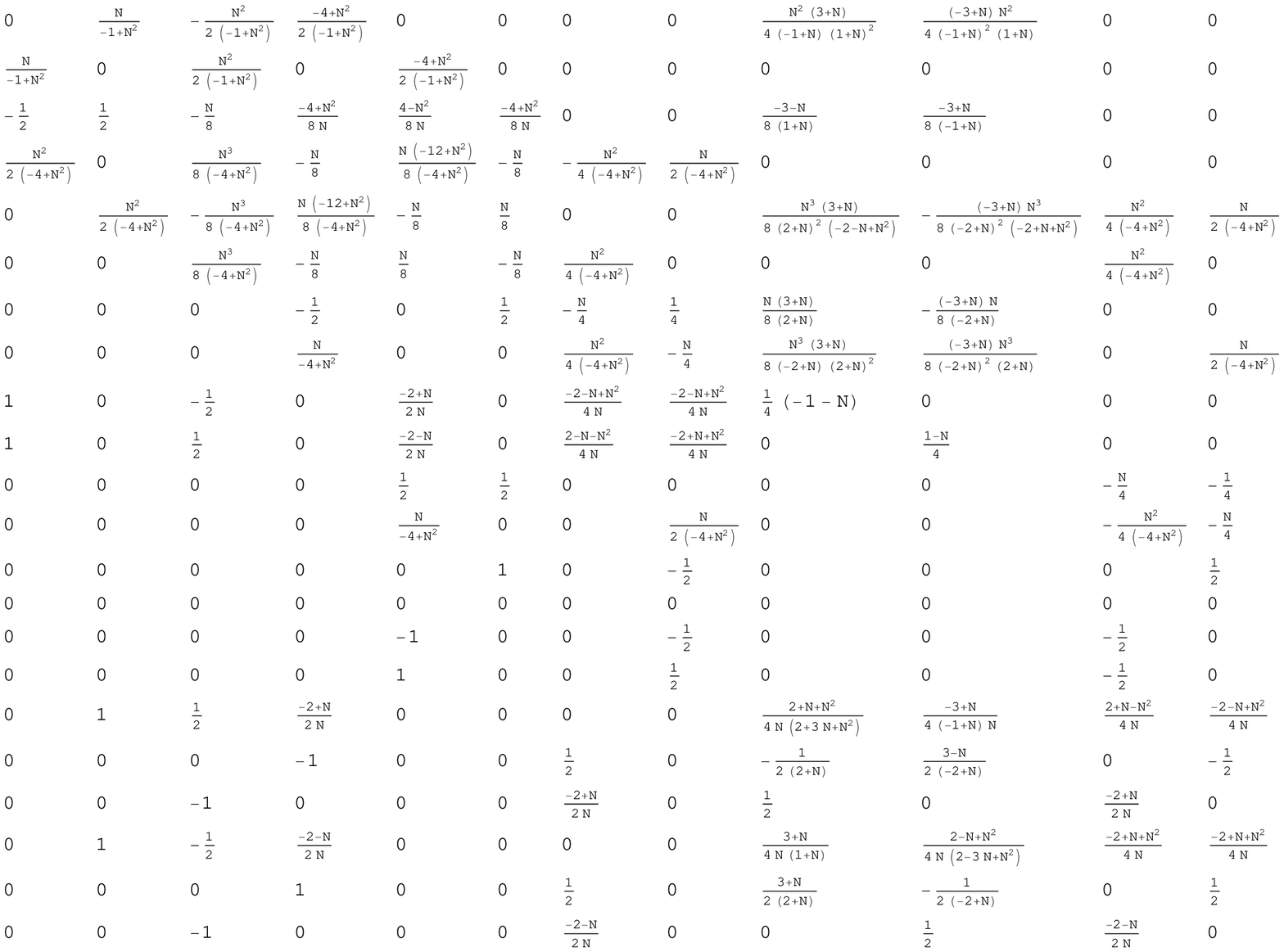}
\end{figure}


\pagebreak
\begin{figure}[h]
\begin{equation}
\label{eq:GPart23}
\end{equation}
\hspace*{15.5 cm}
$\left. \phantom{\begin{array}{llll}
 & | & \\
 & | & \\
 & | & \\
 & | & \\
 & | & \\
 & | & \\
 & | & \\
 & | & \\
 & | & \\
 & | & \\
 & | & \\
 & | & \\
 & | & \\
 & | & \\
 & | & \\
 & | & \\
 & | & \\
 & | & \\
 & | & \\
 & | & \\
 & | & \\
 \end{array}}\right)$
\vspace*{-11 cm}
\\
    \includegraphics[width=16cm]{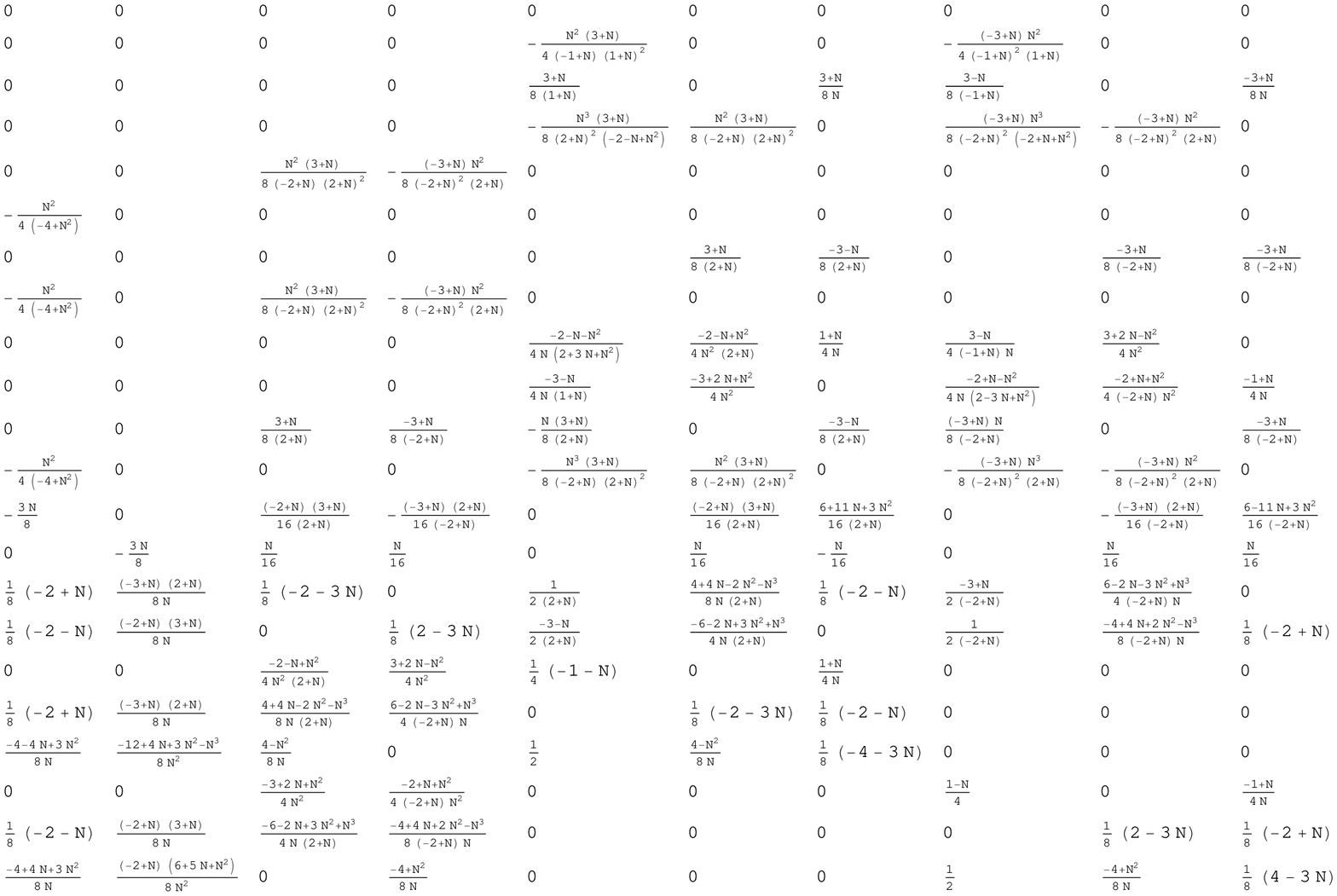}
\begin{equation}
\label{eq:GPart24}
\end{equation}
\hspace*{15.5 cm}
$\left. \phantom{\begin{array}{llll}
 & | & \\
 & | & \\
 & | & \\
 & | & \\
 & | & \\
 & | & \\
 & | & \\
 & | & \\
 & | & \\
 & | & \\
 & | & \\
 & | & \\
 & | & \\
 & | & \\
 & | & \\
 & | & \\
 & | & \\
 & | & \\
 & | & \\
 & | & \\
 & | & \\
 \end{array}}\right)$
\vspace*{-11 cm}
\\
    \includegraphics[width=16cm]{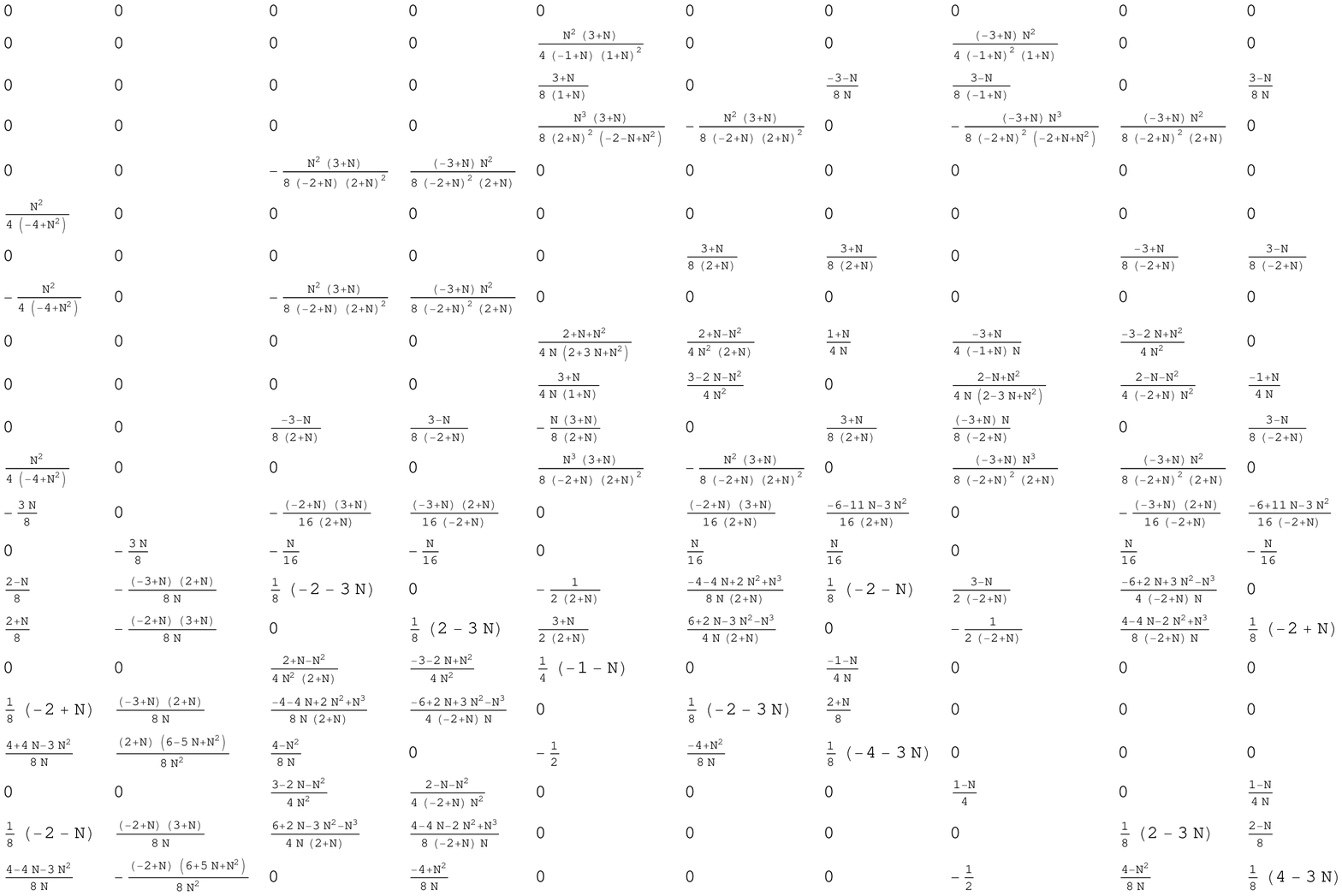}
\end{figure}

\end{document}